\documentclass[smallextended]{svjour3}

\newif\ifdraftmode
\draftmodetrue

\newif\ifshowsplattext

\usepackage[english]{babel}
\usepackage{booktabs}
\usepackage{multirow}
\usepackage{tikz}
\usepackage{xspace}
\usetikzlibrary{matrix,fit,shapes,calc,positioning,shadows,arrows,shapes,backgrounds,decorations.markings,fadings}
\usepackage{listings}
\usepackage[caption=false, font=footnotesize]{subfig}

\usepackage{balance}
\usepackage{wrapfig}
\usepackage{enumitem}
\usepackage{url}

\newcommand{\js}{JS}
\newcommand{\javascript}{JavaScript}
\newcommand{\es}{EcmaScript}

\newcommand{\Comment}[1]{}

\newcommand{\ie}{i.e.}
\newcommand{\eg}{e.g.}
\newif\ifdraftmode
\draftmodetrue
\ifdraftmode
 \newcommand{\Fix}[1]{\textbf{[[}{\color{red} #1}\textbf{]]}}
 \newcommand{\Mar}[1]{\textbf{[[Marcelo: }{\color{magenta} #1}\textbf{]]}}
 \newcommand{\Igor}[1]{\textbf{[[Igor: }{\color{blue} #1}\textbf{]]}}
 \newcommand{\Breno}[1]{\textbf{[[Breno: }{\color{olive} #1}\textbf{]]}}

 \newcommand{\notes}[1]{\todo[inline,color=red!30,caption={}]{#1}}
\else
 \newcommand{\Fix}[1]{\relax}
 \newcommand{\Mar}[1]{\relax}\\
 \newcommand{\Breno}[1]{\relax}
 \newcommand{\Igor}[1]{\relax}
 \newcommand{\notes}[1]{\relax}
\fi

\newcommand{\codesize}{\small}
\newcommand{\CodeIn}[1]{\mcodeid{#1}}

\def\|#1|{\mathid{#1}}
\newcommand{\mathid}[1]{\ensuremath{\mathit{#1}}}
\def\<#1>{\codeid{#1}}
\newcommand{\codeid}[1]{\ifmmode{\mbox{\codesize\ttfamily{#1}}}\else{\codesize\ttfamily #1}\fi}
\def\<#1>{\mcodeid{#1}}
\newcommand{\mcodeid}[1]{\mbox{\codesize\ttfamily{#1}}}

\newcommand{\totalTestFiles}{52,676} 
\newcommand{\totalTestFilesPassInPar}{7,741} 
\newcommand{\totalTestFilesCompileInAll}{38,209} 
\newcommand{\totalTestFilesPassInAll}{26,203} 
\newcommand{\totalTestFilesForTestTransplantation}{9,485} 

\newcommand{\noBugsTransplantation}{35\xspace} 
\newcommand{\noBugsTransplantationConfirmed}{23\xspace}
\newcommand{\noBugsTransplantationFixed}{19\xspace}

\newcommand{\noBugsDifferentialTesting}{24\xspace} 
\newcommand{\noBugsDifferentialTestingConfirmed}{17\xspace}
\newcommand{\noBugsDifferentialTestingFixed}{10\xspace}

\newcommand{\totalBugsReported}{59\xspace}
\newcommand{\totalBugsConfirmed}{39\xspace}
\newcommand{\totalBugsFixed}{29\xspace}

\newcommand{\noBugsBugsReportedSMonkey}{one\xspace}
\newcommand{\noBugsBugsConfirmedSMonkey}{one\xspace}

\newcommand{\noBugsBugsReportedHermes}{four\xspace}
\newcommand{\noBugsBugsConfirmedHermes}{two\xspace}

\newcommand{\noBugsBugsReportedGoogle}{five}
\newcommand{\noBugsBugsConfirmedGoogle}{four}

\newcommand{\noTransVeightBugsReported}{three}
\newcommand{\noTransVeightBugsFixed}{two}

\newcommand{\noTransUndefined}{204} 
\newcommand{\noTransTimeout}{23} 
\newcommand{\noTransNotImplemented}{54} 
\newcommand{\noTransNonStandard}{122} 
\newcommand{\noTransOther}{174} 
\newcommand{\noTransTPDuplicated}{11} 
\newcommand{\noTransTPBugs}{24} 

\newcommand{\noDiffConfirmed}{16}
\newcommand{\noDiffFixed}{ten}
\newcommand{\noDiffVeight}{2}

\newcommand{\noDiffVeightFixed}{1}

\newcommand{\warningsIteration}{five} 

\ifdraftmode
\newcommand{\anonym}[1]{{\tiny\colorbox{black}{xxx}}}
\else
\newcommand{\anonym}[1]{{\tiny\colorbox{black}{#1}}}
\fi

\newcommand{\radamsa}{radamsa\xspace}
\newcommand{\quickfuzz}{quickfuzz\xspace}

\newcommand{\jsc}{JSC\xspace}
\newcommand{\veight}{V8\xspace}
\newcommand{\chakra}{ChakraCore\xspace}
\newcommand{\smonkey}{SpiderMonkey\xspace}
\newcommand{\jerry}{JerryScript\xspace}
\newcommand{\hermes}{Hermes\xspace}

\newcommand{\babel}{Babel\xspace}

\newcommand{\lo}{lo}
\newcommand{\hi}{hi}

\newcommand{\testsThatFail}{63}
\newcommand{\testsThatFailJSC}{6}
\newcommand{\testsThatFailSM}{57}
\newcommand{\filesAttached}{490}

\newcommand{\filesMining}{1,240}

\newcommand{\percentSuiteTestJSC}{92\%}
\newcommand{\percentSuiteTestVeight}{95\%}
\newcommand{\percentSuiteTestChakra}{75\%}
\newcommand{\percentSuiteTestSM}{93\%}
\newcommand{\percentSuiteTestHermes}{26\%}

\newcommand{\failuresTestTrans}{836}
\newcommand{\failuresTestTransDistictFiles}{612} 
\newcommand{\failuresTestTransPercent}{9.2\%}
\newcommand{\noBugsTransplantationSeverityTwo}{16}
\newcommand{\bugsChakra}{12}
\newcommand{\bugsChakraFixed}{3}

\newcommand{\testOriginal}{31,276} 
\newcommand{\testPassInPar}{-}
\newcommand{\testCompileAll}{29,846} 
\newcommand{\testNoFailAll}{17,639} 

\newcommand{\veightOriginal}{1,084} 
\newcommand{\veightPassInPar}{482} 
\newcommand{\veightCompileAll}{478} 
\newcommand{\veightNoFailAll}{426} 

\newcommand{\smOriginal}{3,122} 
\newcommand{\smPassInPar}{2,155} 
\newcommand{\smCompileAll}{2,103} 
\newcommand{\smNoFailAll}{1,837} 

\newcommand{\jscOriginal}{1,265} 
\newcommand{\jscPassInPar}{1,130} 
\newcommand{\jscCompileAll}{1,122} 
\newcommand{\jscNoFailAll}{1,054} 

\newcommand{\duktapeOriginal}{1,195}
\newcommand{\duktapePassInPar}{1,195}
\newcommand{\duktapeCompileAll}{921}
\newcommand{\duktapeNoFailAll}{915}

\newcommand{\jerryOriginal}{1,951}
\newcommand{\jerryPassInPar}{1,951}
\newcommand{\jerryCompileAll}{1,878}
\newcommand{\jerryNoFailAll}{1,837}

\newcommand{\jsiOriginal}{99}
\newcommand{\jsiPassInPar}{99}
\newcommand{\jsiCompileAll}{63}
\newcommand{\jsiNoFailAll}{63}

\newcommand{\tinyOriginal}{49}
\newcommand{\tinyPassInPar}{49}
\newcommand{\tinyCompileAll}{37}
\newcommand{\tinyNoFailAll}{37}

\newcommand{\hermesOriginal}{1,728}
\newcommand{\hermesPassInPar}{680}
\newcommand{\hermesCompileAll}{661}
\newcommand{\hermesNoFailAll}{632}

\newcommand{\babelOriginal}{9,953}
\newcommand{\babelPassInPar}{-}
\newcommand{\babelCompileAll}{2,198}
\newcommand{\babelNoFailAll}{1,745}

\newcommand{\blogEngineNetOriginal}{954}
\newcommand{\blogEngineNetPassInPar}{-}
\newcommand{\blogEngineNetCompileAll}{24}
\newcommand{\blogEngineNetNoFailAll}{18}

\newcommand{\dataRepo}{{\small\url{https://github.com/damorimRG/entente/}\xspace}}

\journalname{Software Quality Journal}

\begin{document}

\title{Exposing Bugs in JavaScript Engines through Test Transplantation and Differential Testing}
\titlerunning{Exposing Bugs in JavaScript Engines}        

\author{Igor Lima \and Jefferson Silva \and Breno Miranda\footnotemark
  \and Gustavo Pinto \and Marcelo d'Amorim}

\institute{
* Corresponding author: Breno Miranda \and
\email{bafm@cin.ufpe.br}\\ \\
Igor Lima \and \email{isol2@cin.ufpe.br} \\
Jefferson Silva \and \email{jefferson.alves.silva@icen.ufpa.br}\\
Breno Miranda \and \email{bafm@cin.ufpe.br} \\
Gustavo Pinto \and \email{gpinto@ufpa.br} \\
Marcelo d'Amorim \and \email{damorim@cin.ufpe.br} \\
\at Universidade Federal de Pernambuco, Recife, PE, Brazil}

\date{Received: date / Accepted: date}



\maketitle

\begin{abstract}
  \textbf{Context.}
JavaScript is a popular programming language today with several
implementations competing for market dominance. Although a
specification document and a conformance test suite exist to guide
engine development, bugs occur and have important practical
consequences. Implementing correct
engines is challenging because the spec is intentionally
incomplete and evolves frequently.
\textbf{Objective.}
This paper investigates the use of test transplantation and
differential testing for revealing functional bugs in JavaScript
engines.  The former technique runs the regression test suite of a
given engine on another engine.  The latter technique fuzzes existing
inputs and then compares the output produced by different engines with
a differential oracle.
\textbf{Method.}
We conducted experiments with engines from five major players--Apple,
Facebook, Google, Microsoft, and Mozilla--to assess the effectiveness
of test transplantation and differential testing.
\textbf{Results.}
Our results indicate that both techniques revealed several bugs, many of which
confirmed by developers. We reported \noBugsTransplantation{} bugs
with test transplantation (\noBugsTransplantationConfirmed{} of these bugs
confirmed and \noBugsTransplantationFixed{} fixed) and reported
\noBugsDifferentialTesting{} bugs with differential testing
(\noBugsDifferentialTestingConfirmed{} of these confirmed
and \noBugsDifferentialTestingFixed{} fixed). Results indicate that
most of these bugs affected two engines--Apple's
\jsc and Microsoft's \chakra{} (24 and 26 bugs, respectively).
To summarize, our results show that
test transplantation and differential testing
are easy to apply and very effective in
finding bugs in complex software, such as JavaScript engines.

\end{abstract}

\section{Introduction}

JavaScript (\js{}) is one of the most popular programming languages
today~\cite{stackify,redmonk-javascript}, with penetration in various
software development segments including, web, mobile, and, more
recently, the Internet of Things~(IoT)~\cite{simply-technologies}. The
interest of the community for the language encourages constant
improvements in its specification~\cite{ecmas262-spec}. It is natural
to expect that such improvements lead to sensible changes in engine
implementations~\cite{kangax}. Even small changes can have high
practical impact. For example, in October 2014 a new attribute added
to Array objects resulted in the MS Outlook Calendar web app to fail
under Chrome~\cite{array-bug-chromium-issue4247,array-bug-discussion}.

Finding bugs in \js\ engines is an important problem given the range
of applications that could be affected with those bugs. It is also
challenging.  Specifications are intentionally incomplete as to enable
development flexibility. In addition, they evolve frequently to
accommodate the pressing demands from
developers~\cite{ecmas262-spec-repo}. An official conformance test
suite exists for \js~\cite{tc39-github}, but, naturally, many test
scenarios are not covered in the suite. In addition, we noticed that a
significant fraction (5 to 15\%) of the tests in that suite fail
regularly in the most popular engines, reflecting the struggle of developers in keeping
up with the pace of spec evolution (see Table~\ref{tab:test262}).

This work, which is empirical in nature, reports on a study to evaluate
the ability of two testing techniques to expose bugs in JavaScript engines.

\begin{itemize}[topsep=0pt,parsep=0pt,partopsep=2pt,labelwidth=0cm,align=left,itemindent=-0.25cm]
\item \emph{Test transplantation}.
  This technique evaluates the effect of running test files
  written for a given engine in other engines. The intuition is that
  developers design test cases with different objectives in mind. As
  such, replaying these tests in different engines could reveal
  unanticipated problems.

\item \emph{Cross-engine differential testing}.  This technique fuzzes
  existing test inputs~\cite{fuzz-testing-history} and then compares
  the output produced by different engines using a differential
  oracle. The intuition is that interesting inputs can be created from
  existing inputs and multiple engines can be used to address the lack
  of oracles for the newly created inputs.
\end{itemize}

We selected these two techniques because they can take advantage
of existing test suites and reuse them --- in a semi-automated way
as we propose here --- to enhance the coverage of the engine under testing.
Cross-engine differential testing is semi-automated as
developers need to decide if alarms are manifestations of real
bugs.
Test transplantation is also semi-automated
and developer intervention may be required to ensure that a test case from one engine can be performed on another engine without adaptations;
or to verify that alarms are manifestations of real bugs and not the result of an unsupported feature or an obsolete specification.
This study measures the ability of these techniques
in finding bugs and the human cost associated with each technique.


%
\emph{Related Ideas.}~Differential Testing~\cite{Brumley-etal-ss07}
(DT) has been applied in a variety of contexts to find
bugs~\cite{Yang-etal-pldi11,Chen-etal-fse2015,Argyros-etla-ccs16,Chen-etal-pldi16,petsios-etal-sp2017,SivakornAPKJ17,Zhang:2017:ATD:3097368.3097448}.
It has shown to be specially practical in scenarios where the
observation of difference gives a strong signal of a real problem. For
example, Mozilla runs JS files against different configurations of a
given build of their \smonkey\ engine (\eg{}, trying to enable or not
eager JIT compilation\footnote{These files are created with the
  grammar-based fuzzer jsfunfuzz~\cite{jsfunfuzz}. Look for option
  ``compare\_jit'' from funfuzz.}). A positive aspect of the approach
is that it can be fully automated---as only one engine is used, the
outcomes of the test in both configurations are expected to be
identical. The Mozilla team uses this approach since 2002; they have
been able to find over 270 bugs since
then~\cite{jsfunfuzz-at-mozilla}, including security
bugs. Cross-engine differential testing, in contrast, has not been
widely popularized because the reported differences are more unlikely
to be false alarms. In this context, a number of legitimate reasons
exist, other than a bug, for a test execution to manifest discrepancy
(see Tables~\ref{fig:piecharts-transplantation} and
~\ref{tab:false-positives}). As consequence, humans need to inspect
the reports.






\begin{figure}[t]
  \centering
  \subfloat[\label{fig:stacked-engine}Bug reports per engine.]{
    \includegraphics[trim=0 150 0 180,clip,width=\textwidth,scale=0.4]{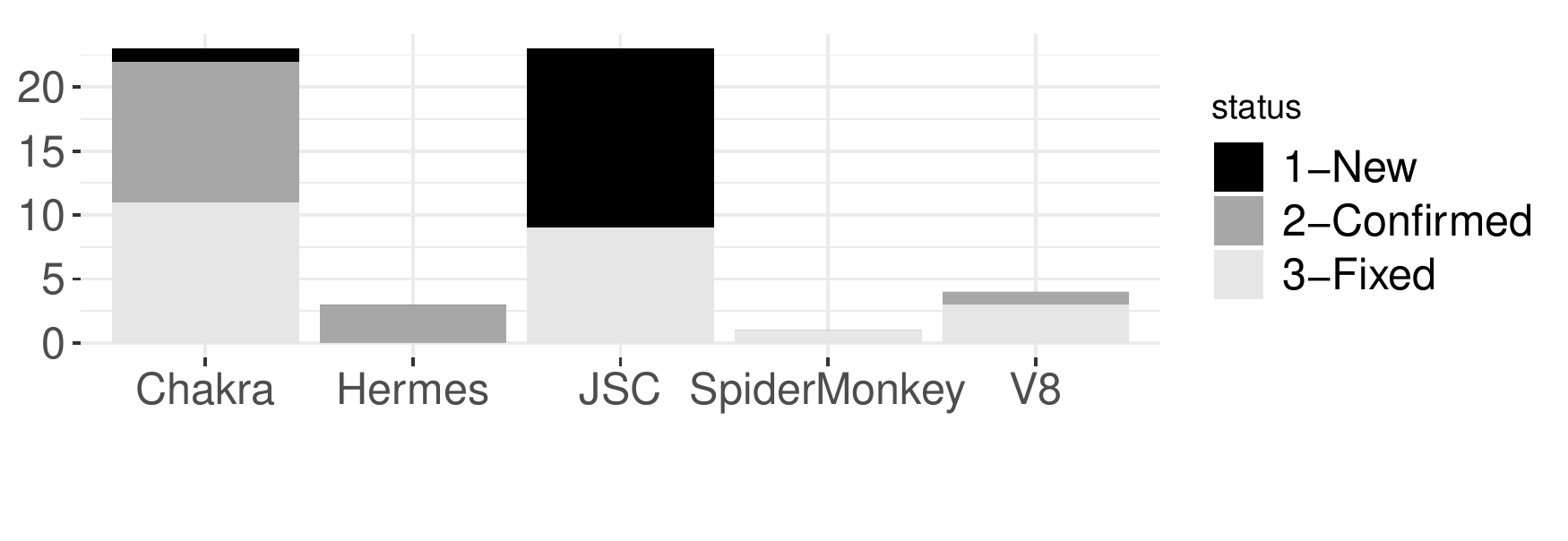}
    \vspace{-5ex}
  }\\
  \subfloat[\label{fig:stacked-technique}Bug reports per technique.]{
    \includegraphics[trim=0 150 0 180,clip,width=0.7\textwidth,scale=0.4]{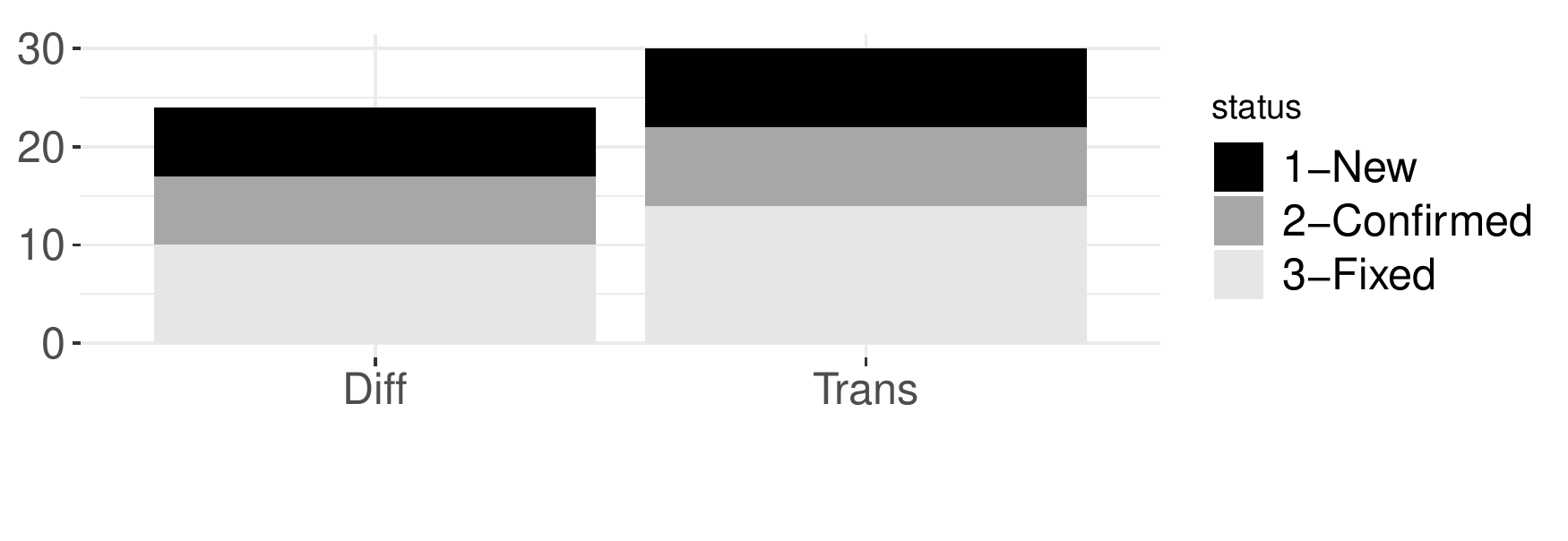}
    \vspace{-5ex}
  }
  \caption{\label{fig:summary}Summary of bug reports.}
  \vspace{-3ex}
\end{figure}

\sloppy \emph{Results.}~We considered the following engines--\chakra{}
(Microsoft), JavaScriptCore (Apple), \veight{} (Google), \smonkey{}
(Mozilla), and \hermes (Facebook). Figure~\ref{fig:summary} shows the
breakdown of bug reports per engine (\ref{fig:stacked-engine}) and per
technique (\ref{fig:stacked-technique}).  Each stacked bar breaks down
the bugs per status (\eg{}, ``1-New''). The prefix number indicates
the ordering that status labels are assigned. Several of these reports
have the label ``3-Fixed'', indicating that bug fixes have been
incorporated into the code already. Note that most of these bugs
affected two engines--\chakra{}\footnote{Microsoft announced in
  December 2018 that the Edge browser will be based on Chromium and
  ChakraCore development would be
  discontinued~\cite{chakra-discontinued}.} and JavaScriptCore (\jsc).
We also reported \noBugsBugsReportedGoogle{} bugs in \veight{}
(\noBugsBugsConfirmedGoogle{} confirmed), \noBugsBugsReportedHermes
bugs in \hermes (\noBugsBugsConfirmedHermes confirmed), and
\noBugsBugsReportedSMonkey bugs in \smonkey{}
(\noBugsBugsConfirmedSMonkey confirmed).  Our results show that both
techniques revealed several bugs, most of which confirmed by
developers. Test transplantation revealed \noBugsTransplantation bugs
(of which, \noBugsTransplantationConfirmed were confirmed and
\noBugsTransplantationFixed were fixed) whereas differential testing
revealed \noBugsDifferentialTesting bugs (of which,
\noBugsDifferentialTestingConfirmed were confirmed and
\noBugsDifferentialTestingFixed were fixed).  Overall, results
indicate that both techniques were successful at finding bugs. The
number of confirmed or fixed bugs are similar. Most bugs we found are
of moderate severity.



\emph{Key Findings.}~To sum up, we found that 1)~Differential testing
and Test Transplantation are practical and effective techniques to
find bugs on real, complex, and widely used software systems and
2)~Even for problems with fairly clear specifications, as in
\javascript{}, there is likely (a lot of) variation between different
implementations, which brings intrinsic challenges to developers that
work on them.  Section~\ref{sec:lessons} expands and elaborates our
key findings and lessons learned.

\emph{Contributions.}~The most important contribution of this work is
empirical: we provide a comprehensive study analyzing the
effectiveness of test transplantation and differential testing in
revealing functional bugs in popular \javascript\ engines.  Additional
contributions include: 1)~A number of bugs found and fixed.  We
reported a total of \totalBugsReported bugs.  Of these,
\totalBugsConfirmed bugs were confirmed and \totalBugsFixed bugs were
fixed.  2)~An infrastructure for performing test transplantation and
differential testing. The source code produced and the generated data sets of tests
and bugs are available on Zenodo~\footnote{https://zenodo.org/record/4116092} and online at the following link: \dataRepo.

\vspace{1ex} To summarize, this paper provides initial, yet strong
evidence that test transplantation and differential testing are simple
and effective techniques to find functional bugs in JavaScript engines
and should be encouraged.

\section{JavaScript}
\label{sec:es6-design}
\label{sec:imp-dep-behavior}

JavaScript engines are virtual machines that parse source code,
compile it in bytecodes, and run these bytecodes. These engines
implement some version of the ECMAScript (\es{}), which emerged with
the goal to standardize variants of the language, such as Netscape's
JavaScript and Microsoft's JScript\footnote{The name JavaScript still
  prevails today, certainly for historical reasons.}. The \es{}
specification is regulated by Ecma International~\cite{es6-website}
under the TC39~\cite{tc39-github} technical committee.  Every year, a
new version of the \es{} specification is released with new features
and minor fixes~\cite{ecmas262-spec-repo,ecmas262-spec}.


The specification of JavaScript is incomplete for different
reasons. Certain parts of the specification are undefined; it is the
responsibility of the community to regulate the evolution of the
language. The JavaScript spec uses the label
``implementation-dependent'' to indicate these cases, where behavior
may differ from engine to engine. One reason for such flexibility in
the spec is to enable compiler optimizations. For example, the
\js\ \mcodeid{for-in} loop construct does not clearly specify the
iteration order of
elements~\cite{so-forin-undefined,javascript-in-chrome} and different
engines capitalize on that for loop
optimizations~\cite{for-in-undefined}.  As another example, the
specification states that if the \mcodeid{Number.toPrecision()}
function is called with multiple arguments then the floating-point
approximation is
implementation-dependent~\cite{es6-toPrecision}. Various other cases
like these exist in the specification. Added to that, given the speed
the specification changes and the complexity of the language some
features are not fully implemented as can be observed by the Kangax
compatibility table~\cite{kangax}.  It is also worth noting that, as
in other languages, some elements in JS have non-deterministic
behavior (\eg{}, \mcodeid{Math.random} and \mcodeid{Date}). A test
that makes decisions based on these elements could, in principle,
produce different outcomes on different runs. Carefully-written test
cases should not manifest this kind of flaky behavior.  As previously
mentioned, all those aspects make testing \js\ engines challenging,
albeit very important given the its tremendous popularity.





\section{Engines Studied}
\label{sec:methodology}
\label{sec:methodology:engines}~We selected
JS engines according to the following criteria: 1) Released latest
version after Jan 1, 2018, 2) Contains more than 1K stars on GitHub,
and 3) Uses a public issue tracker. We looked for highly-maintained
(as per the first criterion) and popular (as per the second criterion)
engines. As we wanted to report bugs, we also looked for project with
public issue trackers. We initially selected four JS engines: \jsc,
\veight, \chakra, and \smonkey. Later, we included \hermes in the list
of studied engines. The main reason was to investigate how our
approach would work on a recently introduced JS engine. More about
\hermes on Section~\ref{sec:hermes}.  Table~\ref{tab:engines} lists
the engines we analyzed. It is worth noting that we used Google Chrome
Lab's JSVU tool~\cite{jsvu} to automatically install and configure
versions of different JS engines in our host environment. This is
important as we aim to use the most recent stable versions of each
engine as to avoid reporting old and already-fixed bugs to developers.


\begin{table}[t]
  \small
  \centering
  \caption{\label{tab:engines}Engines selected.}
  \begin{tabular}{cccrrr}
    \toprule
    Team & Name & URL & \# Stars  & DOB \\
    \midrule
    Apple & \jsc (WebKit) & \cite{jsc2018repo} & 3300+ & Jun 2001 \\
    Google & \veight{} & \cite{v82018repo} & 9800+ & Jun 2008 \\
    Microsoft & \chakra{} & \cite{chakra2018repo} & 7200+ & Nov 2009 \\
    Mozilla & \smonkey{} & \cite{spidermonkey2018repo} & 1100+ & Mar 1996 \\
    Facebook & \hermes & \cite{hermes2020repo} & 5400+ & Jul 2019 \\
   \bottomrule
  \end{tabular}
\end{table}


\section{Mined JS Files}
\label{sec:seeds}

Obtaining good sets of JS test cases is imperative to evaluate the
techniques in this paper. For that, we looked for \js\ files from
various sources: 1) test files from the Test262~\cite{tc39-github}
conformance suite of the ECMA262 specification~\cite{ecmas262-spec},
2) test files from the test suite of our selected engines; these files
are accessible from the engine's official repositories, 3) test files
from the suites of other public engines (\ie{}, Duktape, JerryScript,
JSI, Tiny-js, Babel, and BlogEngine.NET),
and 4) test files mined from issue trackers of these
engines.


\newcommand\marktopleft[1]{%
    \tikz[overlay,remember picture]
        \node (marker-#1-a) at (0,2ex) {};%
}
\newcommand\markbottomright[1]{%
    \tikz[overlay,remember picture]
        \node (marker-#1-b) at (0,0) {};%
    \tikz[overlay,remember picture,thick,dashed,inner sep=2pt]
        \node[draw,rectangle,fit=(marker-#1-a.center) (marker-#1-b.center)] {};%
}

\begin{table}[t]
  \small
  \centering
  \caption{\label{tab:test-suites}Number of test files. Dashed
    rectangle under column ``type-in-all'' shows the tests used for
    test transplantation whereas the rectangle under column
    ``no-fail-in-all''---a subset of the ``type-in-all'' tests---shows
    the tests used in cross-engine differential
    testing.}
  \setlength{\tabcolsep}{2pt}
  \begin{tabular}{ccrrrr}
    \toprule
    \multirow{2}{*}{Name}      &  \multirow{2}{*}{Source} &
    \multicolumn{4}{c}{\# JS files} \\
    \cline{3-6}
                               &         & total & pass-in-par. & type-in-all &  no-fail-in-all \\
    \midrule
    Test262 & \cite{ecma262-conformance-suite} & \testOriginal{} & \testPassInPar{} &  \testCompileAll{} & \marktopleft{c2}\testNoFailAll{} \\
    \midrule
    \jsc & \cite{jsc2018repo} & \jscOriginal{} & \jscPassInPar{} &\marktopleft{c1}\jscCompileAll{} & \jscNoFailAll{}\\
    \smonkey\ & \cite{mozilla} & \smOriginal{} & \smPassInPar{} & \smCompileAll{} & \smNoFailAll{}\\
    \veight{} & \cite{v82018repo} & \veightOriginal{} & \veightPassInPar{} & \veightCompileAll{} & \veightNoFailAll{}\\
    \hermes & \cite{hermes2020repo} & \hermesOriginal{} & \hermesPassInPar{} & \hermesCompileAll{} & \hermesNoFailAll{} \\
    \midrule
    Duktape & \cite{duktape} & \duktapeOriginal{} & \duktapePassInPar{} & \duktapeCompileAll{} & \duktapeNoFailAll{}\\
    \jerry{} & \cite{jerryscript2018repo} & \jerryOriginal{} & \jerryPassInPar{} & \jerryCompileAll{} & \jerryNoFailAll{}\\
    JSI & \cite{jsi} & \jsiOriginal{} & \jsiPassInPar{} & \jsiCompileAll{} & \jsiNoFailAll{}\\
    Tiny-js & \cite{tinyjs} & \tinyOriginal{} & \tinyPassInPar{} & \tinyCompileAll{} & \tinyNoFailAll{}\\
    Babel & \cite{babel2020repo} & \babelOriginal{} & \babelPassInPar{} & \babelCompileAll{} & \babelNoFailAll{} \\
    BlogEngine.NET & \cite{blogEngineNet2020repo} & \blogEngineNetOriginal{} & \blogEngineNetPassInPar{} & \blogEngineNetCompileAll{}\markbottomright{c1} & \blogEngineNetNoFailAll{}\markbottomright{c2} \\
    \midrule
     &  & \totalTestFiles{} & \totalTestFilesPassInPar{} & \totalTestFilesCompileInAll{} & \totalTestFilesPassInAll{}\\
   \bottomrule
  \end{tabular}
\end{table}

Table~\ref{tab:test-suites} shows the breakdown of tests. Column
``Name'' and ``Source'' show the origin of the test suite. Column
``total'' shows the number of test cases associated with a given
source of \js\ files. Column ``pass-in-par.''  shows the number of
test cases that pass in the corresponding engine. We discarded tests
that fail in their engine as we could not reliably indicate the reason
for the failure, so we assumed the test could be broken. We removed
\testsThatFail{} test cases that fail for that
reason--\testsThatFailJSC{} tests from \jsc and \testsThatFailSM{}
tests from \smonkey.

Column ``type-in-all'' shows the number of test cases whose executions
do not throw dynamic type errors in any of the engines because of an
undefined variable or property.  These cases were captured by looking
for the presence of \CodeIn{ReferenceError} and \CodeIn{TypeError} on
the output. A \CodeIn{ReferenceError} (respectively,
\CodeIn{TypeError}) is raised when test execution attempts to access
an undefined variable (respectively, property of an object). We
discarded those tests to avoid noise in the experiments as they
indicate some missing feature in the implementation of the engine as
opposed to bugs. For example, some tests use non-portable names
(\eg{}, \jsc's \CodeIn{drainMicrotasks()} and \smonkey{}'s
\CodeIn{Error.lineNumber}) or use functions that, albeit part of the
spec, not all engines currently support. Similarly, the \chakra
project does not use common assertions available used in JS program,
which other JS engines support. Instead, it uses its own testing
framework, WScript. In such cases, we are unable to reproduce the
\chakra unit tests in other engines---it raises a
\CodeIn{ReferenceError}. Also, updating the other engines to use
WScript would require a non-trivial manual effort. This is why \chakra
does not appear in Table~\ref{tab:test-suites}. For the evaluation of
test transplantation, we used the
\totalTestFilesForTestTransplantation{} tests included in the dashed
rectangle under column ``type-in-all'', \ie{}, all tests under that
column but the Test262 tests. We did not consider tests from the
conformance suite as they are more likely to indicate missing features
as opposed to bugs. In addition, engine developers have access to
these tests and are encouraged to run them.
Column ``no-fail-in-all'' shows the tests for which all engines
pass. The tests in this set are used as fuzzing seeds in the
evaluation of differential testing. The guarantee that tests pass in
all engines assures that discrepancies are related to the changes in
the input produced by fuzzers.


\vspace{0.2cm}
\noindent
\textbf{Cleansing}~We noticed that some of the tests we found
depend on external libraries, which not all selected engines
support. We decided to discard those. For example, we found many tests
based on Node.js~\cite{node} that
require libraries to be installed before running the test and
different tests require different sets of libraries. Supporting those
tests would require an extra setup step and would slow down the
execution of our experiments. Also, as already mentioned, we did not
consider tests from the \chakra{} repository because they depend on
non-portable objects.

\vspace{0.2cm}
\noindent
\textbf{Test Harness}~We noticed that some engines use a
custom shell to run tests, including a harness with specific
assertions. For example, tests provided by Mozilla contain a lot of
custom functions (\eg{}, \CodeIn{assertThrowsInstanceOf},
\CodeIn{assertEqArray}, and \CodeIn{getPromiseResult}) and those are
included in the shell to make the test run correctly.  For that, we
needed to make minor changes in the testing infrastructure to be able
to run the tests uniformly across all engines. More precisely, we
needed to mock non-portable functions, which are only available in
certain engines. Since we had to transplant more than 40k test files,
we tried to mock the minimum amount of code possible to make the test
file work in the other engine. For instance, if a test had a statement
such as \texttt{print(1==1);} we manually refactored that statement to
\texttt{assert(1==1)} which should throw an assertion violation if the
condition returns false.  If we found the change would require
substantial manual work, we opted to discard the test instead.



\vspace{0.2cm}
\noindent
\textbf{Dedup}~The number of tests in \veight\ is low because
we discarded duplicate tests with Mozilla and \jsc. The rationale is to
avoid inflating results and giving credit where it is due. We also
wrote a script that compares each pair of tests from different suites
for similarity. We did not find identical tests, although it is
possible there are equivalent tests modulo renaming.


\subsection{Mining Tests From Issue Trackers}





Test cases embedded in issue trackers are an important source of data
as they may have already shown useful to reveal problems in some
engine. For example, a developer may have found a corner case that a
given engine did not handle properly. Therefore, it is possible that
some other engine does not handle that case as well. For that reasons,
we thought that we should not ignore issue trackers. By manually
analyzing a sample of issues, we observed that developers either 1)
add test cases as attachments of an issue or 2) embed the test cases
within the textual description of an issue. The test cases in
attachments are longer compared to the test cases embedded in issue
descriptions whereas the latter are more common. Consequently, we
thought we should handle both cases.

To obtain test files included as attachments, we wrote a crawler to
visit the issue trackers of all engines listed in
Table~\ref{tab:engines} and we were able to retrieve a total of
\filesAttached{} files.  To mine tests from the textual descriptions we
proceeded as follows. First, we broke the text describing the issue in
paragraphs and used a binary classifier to label each paragraph as
``code'' or ``not code'' (\ie{}, text written in natural language). Then, based on
that information, we merged consecutive paragraphs labeled as ``code''
and used a \js\ parser to check well-formedness of the retrieved code
fragment. Using that method we were able to retrieve a total of
\filesMining{} additional files. All those files were included in
Table~\ref{tab:test-suites}.


For the classification of ``code'' vs. ``nocode'', we used popular
techniques for solving NLP classification
problems~\cite{kusner2015word}.  First, we used
word2vec~\cite{mikolov2013distributed}, a popular NLP technique to
produce word embeddings. A word embedding is a mapping of words to
vectors of real numbers. Then, we used a multi-layer
perceptron~\cite{Rumelhart:1986:LIR:104279.104293} to infer the
probability of the input belonging or not to the class based on the
distance between sentences (\ie{}, the distance from an input sentence
to a code example or to a nocode example) as induced by the distance
of comprising words computed with word2vec. The classifier labels the
input as code if the predicted probability of the input being code is
0.7 or higher.  We used a corpus with 25K samples of English
paragraphs and 25K snippets of \js{} code to train and test the
classifier and obtained an accuracy of 98\%. This classifier is
publicly available from our website as a separate component.

\section{Cross-Engine Differential Testing}
\label{sec:design}

\begin{figure}[t]
  \centering
  \includegraphics[width=0.85\textwidth]{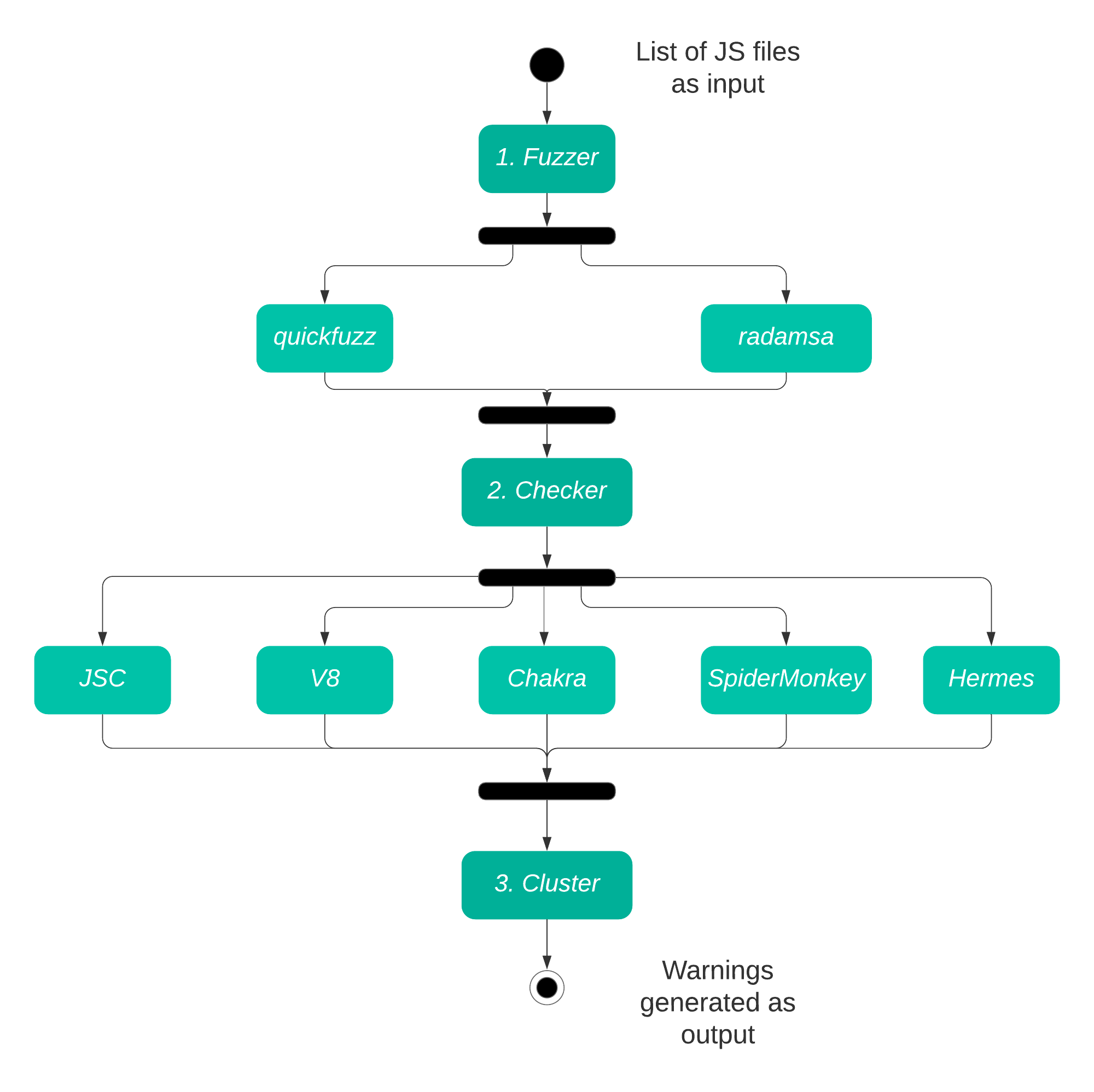}
  \caption{\label{fig:workflow}The workflow of the referred differential
testing approach. In this Activity Diagram, the first black clircle is the start of the process, while the last one if the termination of the process. The green boxes mean activites to be done during the differential testing approach. The small black box mean that there are parallel activities (which in the end are combined).}
\end{figure}

This section describes the infrastructure we used for cross-engine
differential testing. Figure~\ref{fig:workflow} illustrates the
workflow of the approach. It takes on input a list of JS files and
generates warnings on output. Numbered boxes in the figure denote the
data processors and arrowed lines denote data flows. The cycle icons
indicate repetition--the cycle icon close to the cylinder icon
indicates that each file in the input list will be analyzed in
separate whereas the other cycle icon shows that a single file will be
fuzzed multiple times.

The bug-finding process works as follows. First, for a given test
input, the toolchain produces new inputs using some off-the-shelf
input fuzzer (step 1). Section~\ref{sec:objects:fuzzers} describes the
fuzzers we selected. Then, the oracle checks whether or not the output
produced for the fuzzed file is consistent across all engines (step
2). In case the test passes in all engines or fails in all engines
(\ie{}, the output is consistent), the infrastructure ignores the
input. Otherwise, it considers the input as potentially
fault-revealing; hence, interesting for human inspection. Finally, to
facilitate the human inspection process, the infrastructure
prioritizes warnings and clusters them in groups (step 3). We describe
these features in Sections~\ref{sec:prioritization} and
\ref{sec:clusterization}. Note that a number of reasons exist, other
than a bug, for discrepancy to arise (see
Tables~\ref{fig:piecharts-transplantation} and
~\ref{tab:false-positives}) and there is no clear automatic approach
to precisely distinguish false and true positives. As such, a human
needs to inspect a warning to classify the issue. As mentioned
earlier, this justifies why differential testing is challenging to
automate at the functional level. Existing techniques that use
differential testing deal with false alarms differently. For example,
Mozilla varies the configurations of their \smonkey\ engine, but the
implementation is the same~\cite{mozilla-severity}. Non-determinism is
therefore more likely to be associated with the variations across
version as these version use the same core
implementation. CSmith~\cite{Yang:2011:FUB:1993498.1993532} addresses
the problem by trying to avoid generating C files that produce
undefined behavior. Mapping all sources of undefinedness in JS is
impractical.


For step 2, we considered using the open-source tool
eshost-cli~\cite{eshost-cli}, also used at Microsoft, for checking
output discrepancy. However, we noticed that eshost-cli does not
handle discrepancies involving crashes. Our tool supports the execution of the
binaries themselves to obtain these fatal errors.
It is important to note that our checker does not support the case where
the test fails in all engines with a different kind of failure as it
is unclear how to properly relate those failures. Currently, our
infrastructure does not report discrepancy on that case. We left that
as future work as we already found several discrepancies even without
that.



\subsection{Prioritization}
\label{sec:prioritization}

We prioritized warnings based on their likelihood of manifesting a
real bug. We defined two types of warnings based on empirical evidence
we obtained while analyzing bugs. The types ``\hi{}'' and
``\lo{}''. Warnings of the kind ``\hi{}'' are associated with the
cases where the test code executes without violating any internal
checks, but it violates an assertion declared in the test itself or
its harness. The rationale is that the test data is more likely to be
valid in this case as execution does not raise exceptions in
application code. Warnings of kind ``lo'' cover the remaining
cases. These warnings are more likely to be associated with invalid
inputs. They reflect the cases where the anomaly is observed during
the execution of application functions as opposed to assertions. We
observed that different engines often check pre-conditions of
functions differently. It can happen, for example, that one engine
enforces a weaker pre-condition, compared to another engine, on the
inputs of a function and that is acceptable.\Comment{ For instance, it
  is acceptable to pass values greater than \Fix{x} to function
  \CodeIn{\Fix{WWW}} in \Fix{y} but not in \Fix{z}.} In those cases,
the infrastructure would report a warning that is more likely to be
associated with an invalid input produced by the fuzzer, \ie{}, it is
likely to be a ``bug'' in the test code as opposed to a bug in the
engine. Recall that, for differential testing, we only use seed tests
that pass in all engines.

\begin{figure}[h!]
  \centering
  \begin{lstlisting}
var buffer = new ArrayBuffer(64);
var view = new DataView(buffer);
view.setInt8(0,0x80);
assert(view.getInt8(-1770523502845470856) === -0x80);

Message from Engines (1:V8, 2:JavaScriptCore, 3:SpiderMonkey):
1. RangeError: Offset is outside the bounds of the DataView
2. RangeError: byteOffset cannot be negative
3. RangeError: invalid or out-of-range index
  \end{lstlisting}
  \caption{\label{fig:lo-truepositive}Example of a ``\lo'' warning
    that led to a confirmed bug report in \chakra. The bug is caused
    by a required precondition check in the implementation of function
    \CodeIn{ToIndex}~\cite{ecmas262-getviewvalue}, which is indirectly called by the test. }
\end{figure}

Despite the problem mentioned above, ``\lo'' warnings can reveal bugs.
Figure~\ref{fig:lo-truepositive} shows one of these cases. In this
example, the test instantiates an \CodeIn{ArrayBuffer} object and
stores an 8-bit integer at the 0 position. According to the
specification~\cite{ecmas262-getviewvalue}, a \CodeIn{RangeError}
exception should be thrown if a negative value is passed to the
function \CodeIn{ToIndex}, indirectly called by the test case from the
function call \CodeIn{getInt8()}. In this case, however, the \chakra{}
engine did not throw any exception, as can be confirmed from the
report that our infrastructure produces starting with text ``Engine
Messages'' at the bottom of Figure~\ref{fig:lo-truepositive}. This is
a case of undocumented precondition. It was fixed by developers and is
no longer present in the most recent release of the engine.

\subsection{Clusterization}
\label{sec:clusterization}


Clusterization is complementary to prioritization. It helps to group
similar warnings reported by our infrastructure. We only clustered
``\lo'' warnings as ``\hi'' warnings produce messages that arise from
the test case, which are typically distinct.
Figure~\ref{fig:lo-truepositive} shows a test that was originally
available in the \jsc suite. The \radamsa fuzzer mutated the test. It
introduced the negative long \CodeIn{-1770523502845470856} as a
parameter of the \CodeIn{view.getInt8()} method. Before the mutation,
the \CodeIn{view.getInt8()} method received zero as parameter.  At the
bottom of this figure, there is a sequence of three elements that we
use to characterize a warning: 1) the identifier of an engine, 2) the
exception it raises, and 3) the message it produces on a ``\lo''
warning.  This sequence of triples defines a warning signature that we
use for clustering. It is worth mentioning that we filter references
to code in messages as to increase ability to aggregate warnings. Any
warnings, including this one, that has this same signature will be
included in the same ``bucket''. Considering this particular example,
the signature for that cluster will be [(JavaScriptCore,
  ``RangeError'', ``byteOffset cannot be negative''), (SpiderMonkey,
  ``RangeError'', ``invalid or out-of-range index''), (V8,
  ``RangeError'', ``Offset is outside the bounds of the DataView'')].

\subsection{Fuzzers}
\label{sec:objects:fuzzers}


Fuzzers are tools for generating inputs for a given input
format~\cite{8863940}. Different fuzzing strategies exist. We analyzed
\emph{generational} and \emph{mutational} fuzzers.

Generational fuzzers create inputs anew, typically following a
language description, typically context-free grammars. Intuitively,
those fuzzers implement a traversal of the production rules of a
grammar to create syntax trees, which are then pretty-printed and used
as a fresh input. Such strategy to create inputs produces inputs that
are syntactically valid by construction. We analyzed four
grammar-based fuzzers--Grammarinator~\cite{grammarinator},
jsfunfuzz~\cite{jsfunfuzz},
LangFuzz~\cite{Holler:2012:FCF:2362793.2362831}, and
Megadeth~\cite{grieco2016quickfuzz}.  Unfortunately, none of those
were effective out-of-the-box. For example, we produced 100K inputs
with Grammarinator and only few inputs were semantically valid. With
Megadeth, we were able to produce more\Comment{ \Fix{Y, Y$>$Y?}} valid
inputs as it contains some heuristics to circumvent violations of
certain typing rules.\Comment{ such as \Fix{variable used must be
    defined?}.}  Nonetheless, running those inputs in our
infrastructure we were unable to find discrepancies. Inspecting those
inputs, we realized that they reflected very simple scenarios. To sum
up, a high percentage of inputs that Grammarinator and Megadeth
generated were semantically-invalid that we needed to discard whereas
the valid inputs manifested no discrepancies. Considering
jsfunfuzz~\cite{jsfunfuzz}, which has been developed by the Mozilla
team, we noticed that, in addition to the issues mentioned above, it
produces inputs that use functions that are only available in the
\smonkey{} engine. We would need either to mock those functions in
other engines or to discard those tests. Considering
LangFuzz~\cite{Holler:2012:FCF:2362793.2362831}, the tool is not
publicly available. Another fundamental issue associated with
generational fuzzers in our context is that the tests they produce do
not contain assertions; to enable the integration of this kind of
fuzzers in our infrastructure---we would need to look for
discrepancies across compiler error messages as opposed to assertion
violations.  All in all, although grammar-based fuzzers have been
shown effective to find real
bugs~\cite{Holler:2012:FCF:2362793.2362831}, we did not consider those
fuzzers in this study for the reasons above.



Mutational fuzzers modify inputs files provided as seeds. Gray-box
mutational fuzzers use coverage information to guide the mutation
process. American Fuzz Loop (AFL)~\cite{afl} and
libFuzzer~\cite{libfuzzer} are the most popular coverage-guided
fuzzers of today. These fuzzers run tests inputs against instrumented
versions of the programwith the typical goal of finding universal
errors, like crashes and buffer overflows. The instrumentation adds
code to collect branch coverage and to monitor specific
properties\footnote{There are options in the clang toolchain to build
  programs with fuzzing instrumentation~\cite{libfuzzer}. clang
  provides several sanitizers for property
  checking~\cite{clang-documentation}.}. AFL and libFuzzer work very
similarly. They use coverage to find inputs that uncover a new branch
and hence should be fuzzed more. These tools take as input a program
binary (say, a JS engine), which is instrumented to collect coverage
information and to capture runtime violations (\eg{}, illegal memory
accesses), and one seed input to that program (say, a JS program) and
produces on output fault-revealing inputs. We chose to use AFL and
instrument \veight. Unfortunately, we found that most of the inputs
produced by AFL violate the JS grammar. We would need to translate
production rules of Python to the AFL format to circumvent that
issue. We found that the fuzzing task can take days for a single seed
input and there is no clear way to guide the exploration. That happens
because the fuzzer aims to explore the entire decision tree induced
from the engine's main function, including the branches associated
with the higher layers of the compiler (\eg{}, lexer and parser). One
way to mitigate that problem is by writing fuzzing targets
(a.k.a. targets) for specific program functions. Although that
approach has shown to be effective at
Google~\cite{libFuzzer-tutorial-google,libFuzzer-chromium-google}, it
requires domain knowledge to create the calling context to invoke the
fuzz target. For that, we decide not to consider coverage-based in
this study.

We used two black-box mutational fuzzers in this
study: \radamsa~\cite{radamsa} and \quickfuzz~\cite{quickfuzz}. These
fuzzers require no instrumentation and domain knowledge. They mutate
existing inputs randomly. The strength of the approach is
limited by the quality of the test suite and the supported mutation
operators, which are typically simple. We chose these specific fuzzers
because, conceptually, one complements the other. \quickfuzz\ creates
mutations like \radamsa. However, in contrast to \radamsa, \quickfuzz\ is aware
of the \js\ syntax; it is able to replace sub-trees of the syntax
tree~\cite{grieco2016quickfuzz} with trees created anew. Notwithstanding,
since \radamsa performs simpler syntactic source code modifications,
it also produces a higher number of valid inputs.

\section{Results}
\label{sec:results}

The goal of this paper is to assess ability of
test transplantation and differential testing
to find functional bugs
in \javascript\ engines. Based on that, we pose the following three questions:
\begin{description}[leftmargin=.5in]
\item[RQ1.] How conformant are the engines to the Test262 suite?
\item[RQ2.] How effective is test transplantation to find bugs?
\item[RQ3.] How effective is cross-engine differential testing to find bugs?
\end{description}

The first question focuses on the conformance of our selected engines
to the official Test262 suite~\cite{ecma262-conformance-suite}
(Section~\ref{sec:stability}). In the limit, bugs would have low
relevance if the engines are too unreliable. The second question
focuses on the effectiveness of test transplantation
(Section~\ref{sec:transplantation}). The rationale for using inputs
from different engines is that developers consider different goals
when writing tests---suites written for a given engine may cover
scenarios not covered by a different engine. The third question
evaluates the effectiveness of cross-engine differential testing to
find bugs (Section~\ref{sec:cross-engine-diff-testing-results}). The
rationale for this question is that fuzzing inputs may explore
scenarios not well-tested by at least one of the engines.

\subsection{Answering RQ1 (Test262 Conformance)}
\label{sec:stability}

The ECMA Test262~\cite{ecma262-conformance-suite} test suite serves to
check conformance of engines to the \js\ standard. It is
acceptable to release engines fulfilling the specification only
partially~\cite{kangax}. We expect that the pass rate on this suite
provide some indication of the engine's maturity. In the limit, it is
not desirable to flood bug reports on engines at early stages of
development.
For this experiment, we ran the suite once a day for seven consecutive
days and averaged the passing ratios.
We performed seven consecutive executions because,
since the studied JS engines release new versions on a daily basis,
we wanted to make sure if the failing tests raised by the Test262 suite
were rapidly fixed by the maintenance team. If errors were quickly
fixed, this would suggest that a given engine would be closely aligned with
the ECMAScript specification.
Table~\ref{tab:test262} shows
the average number of passing tests over this period. The variance of
results was negligible; for that reason, we omitted standard
deviations. We noticed that all engines but \chakra used some
variant of the Test262 suite as part of their regression process.
We used the same version in this
experiment~\cite{ecma262-conformance-suite}.

\begin{table}
  \centering
  \caption{\label{tab:test262}Percentage of passing tests on
    the Test262 conformance suite.}
  \begin{tabular}{rr}
    \toprule
    engine & \% passing \\
    \midrule
    \veight{} & \percentSuiteTestVeight{} \\
    \smonkey{} & \percentSuiteTestSM{} \\
    \jsc & \percentSuiteTestJSC{}\\
    \chakra{} & \percentSuiteTestChakra{} \\
    \hermes & \percentSuiteTestHermes{} \\
    \bottomrule
  \end{tabular}
\end{table}


Results show that there are still many unsupported scenarios as can be observed from the
percentages in the table. The number of passing tests is high and
similar for \jsc, \veight, and \smonkey.
Moreover, one can also note that \hermes and \chakra have a low passing ratio
in this test suite. Interestingly, \chakra is also the one we were able to find more bugs (as per
Figure~\ref{fig:summary}). Although it is plausible to find
correlation between the passing ratios and reliability as measured by
the number of bugs found, we do not imply causality. As discussed above,
it is important
to note that failures in this conformance test suite indicates missing
features as opposed to bugs. Finally, since \hermes has a
very low adherence to the Test262 conformance suite, we opted to conduct
a case study with them. Therefore, we will only discuss \hermes data at
Section~\ref{sec:hermes}.

\begin{center}
  \fbox{
    \begin{minipage}{11cm}
      \textit{Summary:}~Most of the engines seem to adhere well to the
      \js\ standard. Except for \hermes and \chakra, the passing ratio of all
      engines is above 90\%.
      \end{minipage}
    }
\end{center}

\subsection{Answering RQ2 (Test Transplantation)}
\label{sec:transplantation}

This section reports results of test transplantation. More
specifically, we analyzed the failures observed when running a test
suite original from a given engine in another engine. Intuitively, we
want to assess how effective is the idea of cross-fertilization of
testing knowledge among \js\ developers.

\subsubsection{Methodology}
\label{sec:methodology}
In this experiment, a developer with experience in \js\ analyzed each
test failure, affecting a particular engine, and classified that
failure as potentially fault-revealing or not. The authors supervised
the classification process to validate correctness. For the
potentially fault-revealing cases, one of the authors inspected the
scenario and, if agreed on the classification, reported the bug to the
issue tracker of the affected engine.

\begin{table}[b]
  \small
  \centering
  \caption{\label{tab:cross-testing}Number of failures with
    Test Transplantation.}
  \renewcommand*{\arraystretch}{0.9}
  \begin{tabular}{crrrr}
    \toprule
    test suite\textbackslash{}engine & \jsc & \veight{} & \smonkey{} & \chakra{} \\
    \midrule
    \Comment{
      Lembrar dos testes que os testes da propria engine falham:
      V8 0
      JSC 2
      Spidermonkey 58
    }
    \jsc & - & 10 & 10 & 59 \\
    \veight{} & 41 & - & 3 & 5 \\
    \smonkey{} & 218 & 107 & - & 281 \\
    Duktape & 0 & 4 & 4 & 1 \\
    \jerry{} & 23 & 25 & 22 & 23 \\
    JSI & 0 & 0 & 0 & 0 \\
   Tiny-js & 0 & 0 & 0 & 0 \\
    \midrule
   \textbf{total} & 282 & 146 & 39 & 369  \\
    \bottomrule
  \end{tabular}
  \vspace{-3ex}
\end{table}

\subsubsection{Results}
\label{sec:results}

Table~\ref{tab:cross-testing} shows the number of failures observed
for each pair of test suite and engine. The first column shows the
test suites and the first row shows the engines that run those
tests. We use a dash (``-'') to indicate that we did not consider the
combinations that run the test suite of an engine in itself.  Failures
in those cases would either indicate regressions or flaky tests as
opposed to unknown bugs for that engine. As explained in
Section~\ref{sec:seeds}, we used a total of
\totalTestFilesForTestTransplantation{} tests in this
experiment. These tests are included in the dashed rectangle under
column ``type-in-all'' on Table~\ref{tab:test-suites}. Running those
tests we observed a total of \failuresTestTrans{} failures manifested
across \failuresTestTransDistictFiles{} distinct files
(\failuresTestTransPercent{} of total).  Table~\ref{tab:cross-testing}
shows that \smonkey\ was the engine that failed the least whereas
\chakra\ was the engine that failed the most. The \smonkey\ test suite
also revealed more failures than any other, perhaps as expected, given
that it is the suite with more tests (see
Table~\ref{tab:test-suites}).

In particular, we were not able to reuse the \smonkey tests on the
\hermes engine. This happened because \smonkey has its own assertion
framework, but \hermes did not interpret these assertions, resulting
in failures in all test executions from this engine. These failures
did not happen when the \smonkey tests were transplanted to the other engines,
though.


\sloppy The sources of false positives found in this experiment are as
follows:

\begin{description}
  \item[\textbf{Undefined Behavior.}] False positives of this kind are
  manifested when tests cover implementation-dependent behavior, as
  defined in the ECMA262 specification~\cite{ecmas262-spec}. For
  example, one of the tests from \jerry\ uses the function
  \CodeIn{Number.toPrecision([precision])}, which translates a number to
  a string, considering a given number of significant digits. The
  floating-point approximation of the real value is
  implementation-dependent, making that test to pass only in
  \chakra.
  \item[\textbf{Timeout/OME\footnote{ome is for out of memory
      error.}.}] False positives of this kind typically manifest when the
     engine that runs the test does not optimize the code as the original
     engine of the test. As result, the test fails to finish at the
     specified time budget or it exceeds the memory budget. For example, a
     test case from \jsc defines a function with a tail-call
     recursion. The test fails in all engines but \jsc, which implements
     tail-call optimization.

  \item[\textbf{Not implemented.}] False positives of
this kind manifest when a test fails because it covers a function that
is part of the official spec, but is not implemented in the target
engine yet. For example, at the time of writing, \chakra{} did not
implement by default various properties from the \mcodeid{Symbol}
object. These properties are only available activating the ES6
experimental mode with the flag \CodeIn{-ES6Experimental}.

\item[\textbf{Non-Standard Element.}] These cases manifest when a function or
an object property is undefined in the execution engine but we were
unable to capture that by looking for error types like
\CodeIn{ReferenceError} and \CodeIn{TypeError}.\Comment{ For example,
  we found cases where the test (or its harness) fails in an assertion
  because of an undefined property.}

\item[\textbf{Other.}] This category
includes other sources of false positives. For example, it includes
the cases where the test was valid for some previous version of the
spec but is no longer valid for the current spec.

\end{description}

\begin{table}[h!]
  \centering
  \caption{\label{fig:falsepositives}\label{fig:truepositives}\label{fig:piecharts-transplantation}Distribution
    of False (FP) and True Positives (TP).}
  \renewcommand*{\arraystretch}{0.9}
  \begin{tabular}{ccr}
    \toprule
    & source &  \#\\
    \midrule
    \multirow{5}{*}{FP} & Undefined Behavior & \noTransUndefined{} \\
    & Timeout/OME & \noTransTimeout{} \\
    & Not Implemented & \noTransNotImplemented{} \\
    & Non-Standard Element & \noTransNonStandard{} \\
    & Other & \noTransOther{} \\
    \midrule
    \multirow{2}{*}{TP} & Duplicate & \noTransTPDuplicated{} \\
    & Bug & \noTransTPBugs{} \\
    \bottomrule
  \end{tabular}
\end{table}

Table~\ref{fig:piecharts-transplantation} shows the distribution of
False Positives (FPs) and True Positives (TPs). The sum of the numbers
in this table corresponds to the number of files that manifested
failures, \ie{}, \failuresTestTransDistictFiles{}. Considering false
positives, ``Undefined Behavior'' was the most predominant
source. Considering true positives, we found a reasonable number of
duplicate reports, but not high enough to justify attempting to
automate the detection of duplicates.

\begin{table}[t!]
  \renewcommand{\arraystretch}{0.9}
      \centering
      \caption{List of bugs reports from Test Transplantation.}
      \label{tab:test-transplantation-bugs}

      \begin{tabular}{rccccc}

        \toprule \# & Engine  & Version & Status & Severity & Suite \\
        \midrule
       1  & \jsc{} & 606.1.9.4 & New  & - & \jerry{} \\
       2  & \chakra{}  & 1.9 & \textbf{Confirmed} & 2 & \smonkey{} \\
       3  & \chakra{}  & 1.9 & \textbf{Fixed}   & 2 & \smonkey{} \\
       4  & \chakra{} & 1.10-beta & \textbf{Confirmed} & 2 & \smonkey{} \\
       5  & \jsc{} & 606.1.9.4 & New &  -  & \smonkey{}\\
       6  & \jsc{} & 606.1.9.4 & New & - & \smonkey{} \\
       7  & \jsc{} & 606.1.9.4 & \textbf{Fixed} & 2 & \smonkey{}\\ 
       8  & \chakra{} & 1.10-beta & \textbf{Fixed} & 3 & \smonkey{}\\
       9  & \chakra{} & 1.10-beta & \textbf{Fixed} & 2 & \jsc{}\\
       10 & \chakra{} & 1.10-beta & \textbf{Fixed} & 2 & \smonkey{}\\
       11 & \chakra{} & 1.11-beta & \textbf{Fixed} & 2 & \jsc{}\\
       12 & \chakra{} & 1.11-beta & \textbf{Confirmed} & 2 & \jerry{}\\ 
       13 & \chakra{} & 1.10.1 & \textbf{Fixed} & 2 & \smonkey{}\\
       14 & \jsc{} & 233840 & Duplicated & 2 & \jerry{}\\
       15 & \chakra{} & 1.10.1 & \textbf{Fixed} & 2 & \jerry{}\\ 
       16 & \chakra{} & 1.10.1 & \textbf{Fixed} & 3 & \jerry{}\\
       17 & \jsc{} & 234555 &\textbf{Fixed} & 2 & \jerry{}\\
       18 & \chakra{} & 1.10.1 & \textbf{Fixed} & 3 & \jerry{}\\
       19 & \jsc{} & 234654 & \textbf{Fixed} & 2 & \jerry{}\\
       20 & \veight{} & 7.0.181 & \textbf{Fixed} & 3 & \jerry{}\\
       21 & \jsc{} & 234689 & New & - & \jerry{}\\
       22 & \veight{} & 7.0.237 & WontFix & 2 & Duktape\\
       23 & \chakra{} & 1.10.2 & \textbf{Fixed} & 2 & \smonkey{}\\
       24 & \jsc{} & 235121 & \textbf{Fixed} & 2 & \smonkey{}\\
       25 & \jsc{} & 235121 & \textbf{Fixed} & 2 & \smonkey{}\\
       26 & \veight{} & 7.0.244 & \textbf{Fixed} &  2  & \smonkey{}\\
       27 & \chakra{} & 1.10.2 & \textbf{Fixed} &  2  & \smonkey{}\\
       28 & \jsc{} & 235121 & New & - & \smonkey{}\\
       29 & \chakra & 1.11.19 & \textbf{Confirmed}  & - & \babel \\
       30 & \jsc & 262693 & New & - & \babel \\
       31 & \jsc    & 262693 & New & - & \babel \\
       32 & \smonkey & 77.0b9 & \textbf{Fixed} & 3 & \hermes\\
       \bottomrule
      \end{tabular}
      \vspace{-2ex}
\end{table}

Table~\ref{tab:test-transplantation-bugs} lists all bugs we found with
test transplantation. The first column shows the identifier we
assigned to the bug\Comment{, column ``Date'' shows the date the bug
  was reported in the format ``m/dd'',}, column ``Engine'' shows the
affected engine, column ``Status'' shows the status of the bug report
at the time of the writing. The status string appears in bold face for
status ``Confirmed'' or higher, \ie{}, ``Assigned'' and ``Fixed''.
Column ``Severity'' shows the severity of confirmed bugs, and,
finally, column ``Suite'' shows the name of the engine that originated
the test. Considering severity levels, we found that
\jsc~\cite{jsc-severity} and \smonkey{}~\cite{mozilla-severity}
developers use five levels, whereas \chakra{}~\cite{chakra-severity}
and \veight{}~\cite{v8-severity} developers use only three. As usual,
the smallest the number the highest the severity of the bug. We use a
dash (``-'') in place of the severity level for the cases where the
bug report is pending confirmation. Of the \noBugsTransplantation{}
bugs we reported in this experiment,
\noBugsTransplantationConfirmed\ were promoted from the status ``New''
to ``Confirmed''. Of these, \noBugsTransplantationSeverityTwo{} are
severity-2 bugs.  Although we did not find any critical bugs, most of
the bugs are seemingly important as per the categorization given by
engineers. Analyzing the issue tracker of \chakra, we found that
severity-1 bugs are indeed rare. Considering the number of bug reports
confirmed by developers, \chakra{} was the engine with the highest
number: \bugsChakra{}, with \bugsChakraFixed{} bugs fixed. Considering
the remaining engines, \veight{} developers confirmed the
\noTransVeightBugsReported{} bugs we reported, fixing
\noTransVeightBugsFixed{}. Curiously, Google engineers confirmed the
issued bug reports in a few hours. Likewise, we reported only one bug
on Mozilla's \smonkey, which was quickly fixed.  Overall, we found
that development teams of other engines, specially \jsc, took much
longer to analyze bug reports as can be observed in the \jsc\ stacked
bar from Figure~\ref{fig:stacked-engine}. However, once the team
confirmed those bugs they were then quickly fixed.


\begin{center}
  \fbox{
    \begin{minipage}{11cm}
      \textit{Summary:}~Test transplantation was effective at finding
      functional bugs. Although the cost of classifying failures was
      non-negligible, the approach revealed several non-trivial bugs
      in three of the four engines we analyzed.
      \end{minipage}
    }
\end{center}

\subsection{Answering RQ3 (Differential Testing)}
\label{sec:cross-engine-diff-testing-results}

This section reports the results obtained with cross-engine differential
testing.


\subsubsection{Methodology}
The experimental methodology we used is as follows. As explained on
Section~\ref{sec:objects:fuzzers}, we used Radamsa~\cite{radamsa} and
QuickFuzz~\cite{quickfuzz} for fuzzing. To avoid experimental noise,
we only fuzz test files that pass in all engines--a total of
\totalTestFilesPassInAll\ tests satisfy this restriction. Those tests
appear under the column ``no-fail-in-all'' on
Table~\ref{tab:test-suites}.  We want to avoid the scenario where
fuzzing produces a fault-revealing input based on a test that was
already revealing failures on some engine. This decision facilitates
our inspection task; it helps us establish cause-effect relationship
between fuzzing and the observation of discrepancy.  We configured our
infrastructure (see Figure~\ref{fig:workflow}) to produce 20
well-formed fuzzed files per input file, \ie{}, the number of fuzzing
iterations can exceed the number above as we discard generated files
that are syntactically invalid.


\vspace{0.2cm}
\noindent
\textbf{Exploratory Phase} For the first
three months of the study, our inspection process was exploratory.  In
this phase, we wanted to learn whether or not black-box fuzzers could
reveal real bugs and how effective was the \hi{}-\lo\ warning
classification.  We expected the number of warnings to increase
dramatically compared to the previous experiment and, if we realized
that the ratio of bugs from \lo\ warnings was rather low, we could
focus our inspection efforts on \hi\ warnings. To run this experiment,
we trained eight students in analyzing the warnings that our
infrastructure produced. The students were enrolled in a
graduate-level testing class. We listed warnings in a spreadsheet and
requested the students to update an ``owner'' column indicating who
was working on it, but we did not enforce a strict order on the
warnings the students should inspect. Recall from
Section~\ref{sec:clusterization} that we clustered \lo\ warnings in
buckets. For that reason, we only listed one \lo\ warning per
representative class/bucket in the spreadsheet. First, we explained,
through examples, the possible sources of false alarms they could find
and then we asked the students to use the following procedure when
finding a suspicious warning. Analyze the parts of the spec related to
the problem and, if still suspicious, look for potential duplicates on
the bug tracker of the affected engine using related keywords. If none
was reported, indicate in the spreadsheet that that warning is
potentially fault-revealing. We encouraged students to use
lithium~\cite{lithium} to minimize long test cases\Comment{; the tests
  from the Mozilla suite are often longer than others}. A bug report
was filed only after one of the authors reviewed the diagnosis. Each
student found at least one bug using this methodology.

\vspace{0.2cm}
\noindent
\textbf{Non-Exploratory Phase} Results
obtained in the exploratory phase confirmed our expectations that most
of the bugs found during the initial period of investigation were
related to \hi\ warnings. For that reason, we changed our inspection
strategy. This time, some of the co-authors inspected the bugs using a
similar discipline as before. However, the set of warnings inspected
and the order of inspection changed. We restricted our analysis to
\hi\ warnings and, aware that we would be unable to analyze each and
every warning reported, we grouped those warnings per engine,
analyzing each group in a round-robin fashion.  At each iteration, we
analyzed \warningsIteration{} warnings in each group. A warning
belongs to the group of a given engine if only that engine manifests
distinct behavior, \ie{}, it produces a distinct output compared to
others. We separated in a distinct group the warnings for which two
engines diverge. The rationale for this methodology was to give
attention to each engine more uniformly, enabling more fair comparison
across engines.



\vspace{0.5ex}
\subsubsection{Results}~Table~\ref{tab:summary-hi} shows statistics of \hi\ warnings. The table breaks down \hi\ warning by the affected
engine, \ie, the engine manifesting distinct output among those
analyzed. Column ``+1'' shows the cases where more than one engine
disagree on the output. Note from the totals that the ordering of
engines is consistent with the one observed on
Table~\ref{tab:cross-testing}, with \chakra\ and \jsc\ in first and
second places, respectively, in number of warnings.

\begin{table}[t]
  \small
  \setlength{\tabcolsep}{4.5pt}
  \centering
  \caption{\label{tab:summary-hi}Number of \hi\ warning
    reports per engine.}
  \begin{tabular}{crrrrr}
    \toprule
    fuzzer\textbackslash{}engine & \jsc\ & \veight\ & \chakra & \smonkey & +1\\
    \midrule
    \radamsa{} & 151 & 50 & 331 & 94 & 528 \\ 
    \quickfuzz{} & 83 & 63 & 351 & 21 & 403 \\ 
    \midrule
    \textbf{total} & 234 & 113 & 682 & 115 & 931 \\ 
    \bottomrule
  \end{tabular}
\end{table}

Table~\ref{tab:false-positives} shows the distribution of false
positives per source. The sources of imprecision are as defined in
Section~\ref{sec:transplantation} with the addition of two new
sources, which we did not observe before. These new sources are
marked with a ``*'' in the table. The source ``Invalid Input''
indicates that the test input violated some part of the
specification. For example, the test indirectly invoked some function
with unexpected arguments; this happens because fuzzing is not
sensitive to function specifications. Consequently, it can replace
valid with invalid inputs. The source ``Error Message Mismatch''
corresponds to the cases where the fuzzer modifies the assertion
expression (\eg{}, some string expression or regular expression).

\begin{table}[t]
  \small
  \centering
  \caption{\label{tab:false-positives}Distribution of False (FP)
    and True Positives (TP).}
  \begin{tabular}{ccrr}
    \toprule
    & & \radamsa\ & \quickfuzz\ \\
    \midrule
    \multirow{4}{*}{FP} & Undefined Behavior & 42 & 16 \\ 
    & Timeout/OME & 30 & 15 \\ 
    & * Invalid Input & 46 & 55 \\ 
    & * Error Message Mismatch & 41 & 12 \\ 
    \midrule
    \multirow{2}{*}{TP} & Duplicate & 36 & 28\\ 
    & Bug & 16 & 7\\
    \bottomrule
  \end{tabular}
\end{table}



\begin{table}[t]
  \vspace{-3ex}
  \centering
  \caption{List of bugs reports from Differential Testing.}
  \label{tab:bugs}
  \setlength{\tabcolsep}{3pt}
  \renewcommand{\arraystretch}{0.9}
  \begin{tabular}{rccccccc}
    \toprule
    \# \Comment{ & Date} & Fuzzer & Engine & Version & Status \Comment{&
    \multicolumn{1}{c}{Url}}  & Sev. & Priority & Suite \\
    \midrule
    1  \Comment{& 4/12} & radamsa & \chakra{} & 1.9 & \textbf{Fixed}  \Comment{&
    \anonym{\href{https://github.com/Microsoft/\chakra{}Core/issues/4978}{\#4978}}}
    & 2 & \textbf{\lo} & \jsc{} \\
    2  \Comment{& 4/12} & radamsa & \chakra{} & 1.9 & WontFix  \Comment{&
    \anonym{\href{https://github.com/Microsoft/\chakra{}Core/issues/4979}{\#4979}}}
    & - & \hi{} & \jsc{} \\
    3  \Comment{& 4/14} & radamsa & \jsc{} & 606.1.9.4 & New \Comment{&
    \anonym{\href{https://bugs.webkit.org/show\_bug.cgi?id=184629}{\#184629}}
    } & -  & \hi{} & \jsc{}    \\
    4  \Comment{& 4/25} & radamsa & \chakra{} & 1.9 & \textbf{Fixed}     \Comment{&
    \anonym{\href{https://github.com/Microsoft/\chakra{}Core/issues/5038}{\#5038}}}
    & 2 & \hi{} & \jerry{}   \\
    5  \Comment{& 4/29} & radamsa & \jsc{} & 606.1.9.4 & \textbf{Fixed}  \Comment{&
    \anonym{\href{https://bugs.webkit.org/show\_bug.cgi?id=185127}{\#185127}}
    } & 2  & \hi{}  & \jerry{}\\
    6 \Comment{& 4/30}  & radamsa & \chakra{} & 1.10-beta & \textbf{Confirmed} \Comment{&
    \anonym{\href{https://github.com/Microsoft/\chakra{}Core/issues/5076}{\#5076}}}
    & 2 & \hi{} & TinyJS\\
    7                    \Comment{& 4/30}                       &  radamsa   &
    \jsc{} & 606.1.9.4 & New \Comment{&
    \anonym{\href{https://bugs.webkit.org/show\_bug.cgi?id=185156}{\#185156}}}
    & - & \hi{} & TinyJS \\
    8 \Comment{& 5/02} & radamsa & \jsc{} & 606.1.9.4 & \textbf{Fixed} \Comment{&
    \anonym{\href{https://bugs.webkit.org/show\_bug.cgi?id=185197}{\#185197}}}
    & 2 & \textbf{\lo} & \smonkey{} \\
    9 \Comment{& 5/10} & radamsa & \chakra{} & 1.10-beta & \textbf{Confirmed} \Comment{&
    \anonym{\href{https://github.com/Microsoft/\chakra{}Core/issues/5128}{\#5128}}}
    & 3 & \hi{} & \jerry{} \\
    10 \Comment{& 5/17} & radamsa & \chakra{} & 1.10-beta & \textbf{Fixed} \Comment{&
    \anonym{\href{https://github.com/Microsoft/\chakra{}Core/issues/5182}{\#5182}}}
    & 2 & \hi{} & \veight{}\\
    11 \Comment{& 5/24} & radamsa & \jsc{} & 606.1.9.4 & \textbf{Fixed}  \Comment{&
    \anonym{\href{https://bugs.webkit.org/show\_bug.cgi?id=185943}{\#185943}}}
    & 2 & \hi{} & \jsc{}\\
    12 \Comment{& 6/26} & radamsa & \jsc{} & 606.1.9.4 & \textbf{Fixed}  \Comment{&
    \anonym{\href{https://bugs.webkit.org/show_bug.cgi?id=187042}{\#187042}}}
    & 2 & \hi{} & \jerry{}\\
    13 \Comment{& 7/10} & quickfuzz & \jsc{} & 606.1.9.4 & \textbf{Fixed}  \Comment{&
    \anonym{\href{https://bugs.webkit.org/show_bug.cgi?id=187520}{\#187520}}}
    & 2 & \hi{} & \jerry{}\\
    14 \Comment{& 7/10} & quickfuzz & \chakra{} & 1.11-beta & \textbf{Confirmed}  \Comment{&
    \anonym{\href{https://github.com/Microsoft/\chakra{}Core/issues/5443}{\#5443}}}
    & 2 & \hi{} & \jerry{}\\
    15 \Comment{& 8/21}  & radamsa & \chakra{} & 1.10.2 & \textbf{Fixed} \Comment{&
    \anonym{\href{https://github.com/Microsoft/\chakra{}Core/issues/5617}{\#5617}}}
    & 3 & \hi{} & Test262\\
    16 \Comment{& 8/21} & radamsa &
    \veight{} & 7.0.244 & \textbf{Fixed} \Comment{&
    \anonym{\href{https://bugs.chromium.org/p/v8/issues/detail?id=8078}{\#8078}}}
    & 2 & \hi{} & Test262 \\
    17 \Comment{& 8/23} & quickfuzz & \jsc{} & 235121 &  New  \Comment{&
    \anonym{\href{https://bugs.webkit.org/show_bug.cgi?id=188899}{\#188899}}}
    & - & \hi{} & Test262\\
    18 \Comment{& 8/23} & quickfuzz & \veight{}  & 7.0.244 & \textbf{Confirmed}  \Comment{&
    \anonym{\href{https://bugs.chromium.org/p/v8/issues/detail?id=8088}{\#8088}}}
    & 2 & \hi{} & Test262\\
    19 \Comment{& 8/24}  & quickfuzz & \chakra{} & 1.10.2 & \textbf{Confirmed} \Comment{&
    \anonym{\href{https://github.com/Microsoft/\chakra{}Core/issues/5630}{\#5630}}}
    & 2 & \hi{} & Test262\\
    20 \Comment{& 8/24} &  quickfuzz &
    \jsc{} & 235318 &  New \Comment{&
    \anonym{\href{https://bugs.webkit.org/show_bug.cgi?id=188920}{\#188920}}}
    & - & \hi{} & Test262 \\
    21 \Comment{& 8/24} & quickfuzz & \jsc{} & 235318 & New \Comment{&
    \anonym{\href{https://bugs.webkit.org/show\_bug.cgi?id=188930}{\#188930}}}
    & - & \hi{} & Test262\\
    22 \Comment{& 8/21}  & radamsa & \chakra{} & 1.11.6.0 & \textbf{Confirmed} \Comment{&
    \anonym{\href{https://github.com/Microsoft/\chakra{}Core/issues/5968}{\#5968}}}
    & - & \hi{} & \smonkey{}\\
    23 & radamsa & \chakra & 1.11.19 & \textbf{Confirmed} & - & \lo & \hermes \\

   \bottomrule
  \end{tabular}
\end{table}

Table~\ref{tab:bugs} shows the list of bugs we reported. The table
shows the fuzzing tool used
(``Fuzzer''), the \js\ engine affected (``Engine''), the status of the
bug report (``Status'')\Comment{ as of Aug. 24, 2018, the URL of the bug report
(``URL'')}, the severity of the bug report (``Sev.''), the priority
that we assigned to the warning that revealed the bug (``Priority''),
and the test suite from the original test input (``Suite''). So far,
\noDiffConfirmed{} of the bugs we reported were confirmed, \noDiffFixed{} of which
were fixed. Note that one bug report that we submitted was rejected on
the basis that the offending JS file manifested an incompatibility
across engine implementations that was considered to be acceptable. As
of now, we did not find any new bugs on \smonkey{}; the bugs we found
were duplicates and were not reported. For \veight{},
we reported \noDiffVeight{} bugs,
all of them confirmed, with \noDiffVeightFixed{} fixed.



\begin{center}
  \fbox{
    \begin{minipage}{11cm}
      \textit{Summary:}~Cross-engine differential testing was
      effective at finding \js\ engines bugs, several of which have
      been fixed already.
      \end{minipage}
    }
\end{center}

\noindent
\textbf{Data Availability.}~The data, including the tests, warning
reports, and diagnostic outcomes, is publicly available from a
preserved repository \dataRepo.

\section{Case Study: \hermes}
\label{sec:hermes}

This section describes the experiment we conducted to evaluate the
effectiveness of the techniques we studied to find new bugs in
\hermes, a JavaScript engine that has started development very
recently. \hermes is a JavaScript engine introduced by Facebook in
2019. According to its website, \hermes is ``a JavaScript engine
optimized for fast start-up of React Native apps on
Android''~\cite{hermes2020repo}. Currently, \hermes\ is used as a beta
component of the react-native framework, which is also maintained by
Facebook. The goal of \hermes\ at Facebook is to increase the
performance of Android applications, such as startup time, to reduce
memory consumption and application size~\cite{reactnative2020hermes}.

As a preparatory step to apply test transplantation and differential
testing to \hermes, we ran the Test262 conformance suite on the most
recent version of \hermes, 0.5.0. We observed that only 26\% of the
Test262 tests passed. When taking a closer look, we noticed that this
high percentage of failing tests was due to the high number of
features not yet supported by \hermes\footnote{See
  \url{https://github.com/facebook/hermes/blob/master/doc/Features.md}}. For
instance, in its most recent release, \hermes still does not support
for features such as Proxy, Promise, and Async calls. For this
reason---low coverage on Test262---, we opted not to include \hermes
with other JavaScript engines. Since we only fuzz test files that pass
in all engines, this would require us to significantly reduce the
number of tests used in the experiments, affecting overall results.

We followed the same methodology described in previous sections to
evaluate \hermes. When experimenting with test transplantation, we
noted many errors when replaying the test suite of \smonkey on
\hermes. This happened because \smonkey has its own assertion
framework, but \hermes was unable to interpret those assertions
(because of missing features), resulting in failures in all test
executions from. These failures did not happen when the \smonkey tests
were transplanted to the other engines, though.  Overall, we reported
three issues to the \hermes bug tracker as result of using test
transplantation\footnote{\url{https://github.com/facebook/hermes/issues/<id>},
  with id 265, 266, 267.}. One of them was confirmed by the \hermes
maintainers by the time of this submission. For another, the
maintainer explained the issue was related to a feature not yet
supported---a \texttt{return} statement outside the scope of a
block. For the third issue, developers disagreed on the proper
implementation of one particular part of the specification involved in
the issue (i.e., whether a numeric escape character (non-octal-eight,
\texttt{\textbackslash{8}}) should be allowed or not in strict
mode. The issue was closed, but the maintainers opened a new one to
discuss this matter. By following the discussion, it was interesting
to observe that maintainers are aware that this feature is implemented
by other engines such as \veight, \chakra, and \smonkey. Yet, they
were unsure how they should treat it in \hermes.  Considering
differential testing, we found and reported one bug on \hermes using
\radamsa. As of this writing, this particular bug report was not yet
addressed by any \hermes maintainer.

Overall, we noted that test transplantation and differential testing
were effective techniques in revealing potential bugs in non-trivial
JavaScript engines of varying degrees of maturity. We found bugs both
in extremely robust and largely adopted engines and also in recently
developed engines. The issue diagnosis of
developers seem to differ though. Issues reported to stable engines
are more likely to be real bugs---as lots of problems have been
already scrutinized---whereas issues reported to new engines are more
likely to be inconsistencies related to influx development as opposed
to real bugs.

\section{Discussion}


This section discusses bug reports, threats to validity, and key
findings and lessons learned.

\subsection{Bug Reports}

We issued several bug reports as result of this work. For space, we
are unable to discuss all of them. We sampled some bug reports to
discuss in the following. The selection criteria we used was:
\begin{enumerate}
\item to cover all engines we found bugs--\chakra, \hermes, \jsc,
  \smonkey, \veight (see Figure~\ref{fig:summary});
\item to cover each technique--test transplantation and differential
  testing;
\item to cover a case of rejected bug report;
\item to use short tests (for space).
\end{enumerate}


\sloppy

\vspace{1ex}\noindent\textbf{Issue \#19,
  Table~\ref{tab:test-transplantation-bugs}.} The code snippet below
shows the test input we used to reveal a bug in \textbf{\jsc} version 234555.

\begin{figure}[h!]
  \vspace{-0.5ex}
  \centering
  \scriptsize
  \lstset{escapeinside={@}{@},
    basicstyle=\ttfamily\scriptsize, boxpos=c,
    numberstyle=\tiny,
    morekeywords={assertEq, var, yield, in, function,
    typeof, return, throw, new, Error, if},
  }
  \begin{lstlisting}
 var obj = {}; var arr = [];
 try { arr.sort(obj); assert(false);}
 catch (e) { assert(e instanceof TypeError); }
  \end{lstlisting}
  \normalsize
  \vspace{-1ex}
\end{figure}

This is a test case of the \jerry{} suite. The bug was found during
the test transplantation experiment. According to the
\es\ spec~\cite{ecmas262-array-sort}, the parameter to the
\CodeIn{Array.sort} function should be a comparable object or an
undefined value, otherwise it should throw a \CodeIn{TypeError}. In
this case, \jsc incorrectly accepts a non-callable object as argument
to \CodeIn{sort} and the test fails in the subsequent step. The other
engines raise a \CodeIn{TypeError} as expected.

\vspace{1ex}\noindent\textbf{Issue \#32, Table~\ref{tab:test-transplantation-bugs}.} We
reported the snippet \CodeIn{"use strict" 010} to the
\textbf{\smonkey} development team as a bug and the bug was confirmed
in a few hours.  This is a test case originally from the \hermes suite
using test transplantation.  According the specification, in strict
mode, the engine must use the prefix '0o' or '0O' to represent octal
numeric literals. This is an issue that explores a deprecated octal
token after ASI (Automatic Semicolon Insertion). The other engines
throw a SyntaxError due to the missing of the octal token, but \smonkey
returns an integer \CodeIn{8} that represents \CodeIn{010} in octal.


\vspace{1ex}\noindent\textbf{Issue \#2, Table~\ref{tab:bugs}.} We
reported the code snippet below to the \textbf{\chakra} development team as a
bug, but they did \emph{not} accept.

\begin{figure}[h!]
  \vspace{-0.5ex}
  \centering
  \scriptsize
  \lstset{escapeinside={@}{@},
    basicstyle=\ttfamily\scriptsize, boxpos=c,
    numberstyle=\tiny,
    morekeywords={assertEq, var, yield, in, function,
    typeof, return, throw, new, Error, if},
  }
  \begin{lstlisting}
 function test() {
   return typeof String.prototype.repeat === "function"
     && "foo".repeat(657604378) === "foofoofoo"; }
  \end{lstlisting}
  \normalsize
  \vspace{-1ex}
\end{figure}

This is a test case original from the \jsc suite that the
\radamsa\ fuzzer modified. The original test used the integer literal
3 as argument to \CodeIn{repeat()}, \ie{}, the expression produced a
string with three repetitions of the string ``foo''. The new test uses
a long integer instead as parameter to \CodeIn{repeat()}. As result,
the engine crashes. The team answered that this was an incompatibility
by design\footnote{We interpreted as a violation of an undocumented
  precondition} as the function was not expected to receive such a
long value.

\vspace{1ex}\noindent\textbf{Issue \#4, Table~\ref{tab:bugs}.} The
code snippet below shows a test that reveals a bug in \textbf{\chakra}.

\begin{figure}[h!]
  \vspace{-1ex}
  \centering
  \scriptsize
  \lstset{escapeinside={@}{@},
    basicstyle=\ttfamily\scriptsize, boxpos=c,
    numberstyle=\tiny,
    morekeywords={assertEq, var, yield, in, function,
    typeof, return, throw, new, Error, if},
  }
  \begin{lstlisting}
 { var a = {valueOf: function(){ return "\x00"}}
   assert(+a === 0) }
  \end{lstlisting}
  \normalsize
  \vspace{-1ex}
\end{figure}

The object property \CodeIn{valueOf} stores a function that returns a
primitive value identifying the target object~\cite{valueof}. The
original version of this code returns an empty string whereas the
version of the code modified by the \radamsa{} fuzzer~\cite{radamsa}
returns a string representation of a null character
(\CodeIn{NUL})\Comment{ in ascii}. The unary plus expression
``\CodeIn{+a}", used in the assertion, is equivalent to the operation
\CodeIn{ToNumber(a.valueOf())} that converts a string to a number,
otherwise the operation returns NaN (Not a
Number)~\cite{unary-plus}. This test fails in all engines but
\chakra{}. For all three engines the string cannot be parsed as an
hexadecimal. As such, they produce a NaN and the test fails as
expected. \chakra{}, instead, incorrectly converts the string to zero,
and the test passes. As Table~\ref{tab:bugs} shows, the \chakra{} team
fixed the issue soon after our report.

\vspace{1ex}\noindent\textbf{Issue \#18, Table~\ref{tab:bugs}.} The
snippet {\footnotesize\ttfamily{eval(function b(a)\{break;\});}}
revealed a bug in \textbf{\veight{}} version 7.0.244. This code snippet was
obtained by fuzzing a Test262 test with \quickfuzz. In its original
version, a string (omitted for space), passed as argument to the
\CodeIn{eval} function, encoded the actual test. The fuzzer replaced
the string argument with a function whose body is a \CodeIn{break}
statement outside a valid block statement. Section B.3.3.3 from the
\es\ spec~\cite{spec-b333} documents how eval should handle code
containing function declarations.  According to the
spec~\cite{break-statement}, the virtual machine should throw an early
error--in this case, a SyntaxError--if the break statement is not
nested in a loop or switch statement. All engines, but \veight{},
behave as expected in this case.


\vspace{1ex}\noindent\textbf{Bug reported on \hermes.} The snippet
\CodeIn{b+/v/a} represents a bug confirmed in \textbf{\hermes} engine with
differential testing using radamsa.
The original test case contains a string concatenation.
This is a case of validation of RegEx flags. The fuzzer mutates
the file with a regular expression \CodeIn{/v/} after the plus operator.
In this case, the plus operator turns into a flag of the RegEx \CodeIn{+/v/},
but this flag is not valid. The expected behavior is to throw an early
SyntaxError due to the invalid regular expression flag,
but \hermes\ seems not to treat the early validation of the regexp flags
as explained by one of their developers.

\subsection{Threats To Validity}

As it is the case of most empirical evaluations, our findings are
subject to internal, external, and construct threats to
validity. Considering internal validity, conceptually, it is possible
that the authors of this paper made mistakes in the implementation of
the scripts supporting the experiments. To mitigate this threat, we
carefully inspected the implementation and results, looking for
inconsistencies whenever possible. As for external validity, our
results might not generalize to other test inputs and engines.
We though carefully selected inputs from various
sources according to a well-defined criteria (Section~\ref{sec:seeds}).
Likewise, we selected
the engines by using using a documented criteria (Section~\ref{sec:methodology:engines})
and found that the
engines selected were associated, certainly not by coincidence, with
the browsers informally considered the most popular in the market.  A
reader might argue that since we mined files from other JS engines,
such as Duktape, JerryScript, JSI, and Tiny-js, we could also have ran
test transplantation on them. We decided not to do so for different
reasons: JSI was deprecated, TinyJS is not actively
maintained. Finally, at the time we started this work, Duktape and
JerryScript did not provide support to ES6. We checked this
information again in Jun 2020, and they are still only partially
supporting ES6. In terms of construct validity, we used standard
metrics to determine the effectiveness of the testing techniques we
studied (\eg{}, number of bugs confirmed and fixed and
severity). Engine developers were responsible for determining the
labels of the bug reports and their severity. Consequently, these
metrics originate from a trusted source.


\subsection{Key Findings}
\label{sec:findings}

The main findings of this study are as follows.

\begin{enumerate}
  \item Both techniques we studied have shown to be practical and
    effective to find bugs on real, complex, and widely used software
    systems;
  \item Even for language APIs with fairly clear specs, as it is the
    case of JavaScript, there is likely (a lot of) variation between
    different implementations, which brings intrinsic challenges to
    developers that work on them;
  \item Even simple black-box fuzzers can create surprisingly
    interesting inputs;

\end{enumerate}


The main findings of this study are as follows: 1) The techniques we
selected found, with relatively low effort, several bugs even in very
robust engines, such as Mozilla's \smonkey, 2)~Even for software
projects with fairly clear specifications, as the case of
\javascript{}~\cite{ecmas262-spec}, there are lots of undefined
implementation-specific behaviors whose implementation can
inadvertently lead to interference in the well-specified parts of the
spec, leading to clear bugs. 3)~Finding functional/non-crash bugs with
differential testing is feasible on real, complex, widely used pieces
of software. Even black-box mutational fuzzers revealed bugs. We do
expected that other kinds of fuzzers (\eg{}, graybox fuzzers and
black-box grammar-based generational fuzzers) could reveal even more
bugs

\subsection{Key Lessons}
\label{sec:lessons}

The key lessons of this study are as follows:

\begin{enumerate}
\item The use of Mozilla's \smonkey\ and Google's \veight\ engines
  should be encouraged;
\item The cost of inspection of the warnings in differential testing
  should be reduced;
\item Finding bugs in large software is a very effective way to engage
  students in contributing to open-source software and learning
  Software Testing practices.
\end{enumerate}

1)~When taking into consideration our bug finding campaign, Mozilla's
\smonkey\ and Google's \veight\ stood out as the most reliable
engines. Therefore, we recommend the use of these engine for running
\js\ applications. 2)~Further reducing cost of inspection in
differential testing is an important problem. Although the inspection
activity was not uninterrupted, it is safe to say that each warning
required a substantial amount of time to analyze for potential false
alarms. In fact, many \hi\ warnings reported with differential testing
were not analyzed. We observed empirically that the cost of analysis
was proportional to (i) the \js\ specification covered by the original
test (as developers need to read and understand those parts) and (ii)
the availability of alternative implementations. We prefer to see such
problem as an opportunity for future research. For example, applying
learning techniques to prioritize the warnings more likely to be
faulty (in the spirit of the work of Chen and
colleagues~\cite{Chen:2017:LPT:3097368.3097451}) may be a promising
avenue to explore. Recall that the rate of true positives of the
techniques we studied is rather small. 3)~We learned that reporting
real bugs is a great way to train (and encourage) students in software
testing. Students praised the experience of diagnosing failures,
understanding part of the specs (as needed), writing bug reports,
participating in discussions on issue trackers, and observing the
change of status. That was a relatively self-contained hands-on
activity that enabled students to engage in a real-life serious
industrial project.

\section{Related Work}
\label{sec:related-work}


\subsection{Differential Testing}
Several different applications of differential testing have been
proposed in recent years. Chen and
colleagues~\cite{Chen:2018:RDT:3180155.3180226} recently proposed a
technique to generate X.509 certificates based on Request For
Proposals (RFC) as specification with the goal of detecting bugs in
different SSL/TLS implementations. Those bugs can compromise security
of servers which rely on these certificates to properly authenticate
the parties involved in a communication session. Lidbury and
colleagues~\cite{Lidbury:2015:MCF:2737924.2737986} and Donaldson and
colleagues~\cite{Donaldson:2017:ATG:3152284.3133917} have been
focusing on finding bugs in programs for graphic cards (\eg{},
OpenCL). These programs use the Single-Instruction Multiple-Data
(SIMD) programming abstraction and typically run on GPUs.  Perhaps the
application of differential testing that received most attention to
date was compiler testing. In 1972, Purdom~\cite{Purdom1972} proposed
the use of a generator of sentences from grammars to test correctness
of automatically generated parsers. After that, significant progress
has been made. Lammel and Shulte proposed Geno to cross-check XPath
implementations using grammar-based testing with controllable
combinatorial coverage~\cite{10.1007/11754008_2}. Yang and
colleagues~\cite{Yang:2011:FUB:1993498.1993532} proposed CSmith to
randomly create C programs from a grammar, for a subset of C, and then
check the output of these programs in different compilers (\eg{}, GCC
and LLVM). Le and colleagues~\cite{Le:2014:CVV:2594291.2594334}
proposed ``equivalence modulo inputs'', which creates variants of
program which should have equivalent behavior compared to the
original, but for which the compiler manifests
discrepancy. Differential testing has also been applied to test
refactoring engines~\cite{Daniel:2007:ATR:1287624.1287651}, to test
symbolic engine implementations~\cite{Kapus:2017:ATS:3155562.3155636},
to test disassemblers and binary
lifters~\cite{Paleari:2010:NDD:1831708.1831741,Kim:2017:TIR:3155562.3155609},
and very recently to test JavaScript
debuggers~\cite{DBLP:conf/sigsoft/LehmannP18}. All in all, it has
shown to be flexible and effective for a wide range of
applications. Surprisingly, not much work has been done on
differential testing of \js\ engines. Mozilla uses differential
testing to look for discrepancies across different configurations of
the same version of its \smonkey\ engine (using the ``compare\_jit''
flag of jsfunfuzz~\cite{jsfunfuzz}) whereas we focus on discrepancy
across engines. Patra and Pradel evaluated their language-agnostic
fuzzing strategy using differential testing. Their focuses on finding
differential bugs across multiple
browsers~\cite{patra2016learning}. As such they specialized their
fuzzer to HTML and JS (see Section~\ref{sec:testing-js-engines}). In
contrast to Patra and Pradel, we did not propose new techniques; our
contribution was empirical.

\subsection{Testing \js\ Programs}
Patra and colleagues~\cite{Patra:2018:CFU:3180155.3180184} proposed a
lightweight approach to detect conflicts in \js\ libraries that occur
when names introduced by different libraries collide. This problem was
found to be common as the design of \js\ allows for overlaps in
namespaces. A similar problem has been investigated by Nguyen and
colleagues~\cite{nguyen-etal-icse2014} and Eshkevari and
colleagues~\cite{eshkevari-etal-icpc2014} in the context of PHP
programs, which are popular in the context of Content Management
Systems as WordPress. The focus of this paper is on testing
\js\ engines as opposed to \js\ programs. Our goal is therefore
orthogonal to theirs.

\subsection{Testing \js{} Engines}
\label{sec:testing-js-engines}
The closest work to ours was done by Patra and
Pradel~\cite{patra2016learning}. Their work proposes a
language-agnostic fuzzer to find cross-browser HTML+JS
discrepancies.\Comment{ They applied the proposed fuzzer to find bugs in JS engines and
Web Assembly engines.} The sensible parts of the infrastructure they
built are the checks of input validity (as to reduce waste/cost) and
output correctness (as to reduce false positives). Patra and Pradel
work is complementary to ours--in principle, we could use their fuzzer
in our evaluation. The main difference of our work to theirs is in
goal--we aim at assessing reliability of \js\ engines and find bugs on
them using simple approaches whereas they aim at proposing a new
technique.

Fuzzing is an active area of investigation with development of new
techniques both in academia and industry. Several fuzzing tools exist
focused on \js. Section~\ref{sec:objects:fuzzers} briefly explain
different fuzzing strategies and tools. Existing techniques prioritize
automation with a focus on finding crashes; see the sanitizers used in
libFuzzer~\cite{libfuzzer-tutorial}, for instance. In general, it is
important for these tools that a warning reveals something potentially
alarming as a crash given that fuzzing is a time-consuming operation,
\ie{}, the ratio of bugs found per inputs generated is often very low.
Our approach contrasts with that aim as we focus on finding errors
manifested on the output, which rarely result in crashes and,
consequently, would go undetected by current fuzzing approaches. It is
should be noted, however, that such problems are not unimportant as
per the severity levels reported in
Tables~\ref{tab:test-transplantation-bugs} and~\ref{tab:bugs}.

\section{Conclusions}


JavaScript (\js{}) is very popular today. Bugs in engine
implementations often affect lots of people and organizations.
Implementing correct engines is challenging because the specification
is intentionally incomplete and evolves frequently. Finding bugs in
\js\ engines is challenging for similar reasons.

This paper reports on a study to evaluate two
techniques for finding bugs in \js{}--test transplantation and
cross-engine differential testing. The first technique
runs the test suite of one given engine in another engine.
The second technique fuzzes existing inputs and then compares the
output produced by different engines with a differential oracle.

We found that both techniques were very effective at finding bugs in
\js\ engines.  Overall, we reported \totalBugsReported\ bugs in this
study. Of which, \totalBugsConfirmed\ were confirmed by developers and
\totalBugsFixed\ were fixed. Although more work is necessary to reduce
cost of manual analysis, we found that our results provide strong
evidence that exploring test transplantation and differential testing
should be encouraged to find functional bugs in JavaScript engines.

In the near future, we plan to explore techniques to prioritize the
warnings reported by these techniques and to continue involving
students in the task of diagnosing these warnings. Using learning
techniques for prioritization~\cite{icst2020violations},
similar to what Chen and colleagues~\cite{Chen:2017:LPT:3097368.3097451}
did to prioritize the warnings reported by CSmith~\cite{Yang:2011:FUB:1993498.1993532},
seems a promising starting point for this improvement.

The scripts to run the experiments for this study will be available
upon request. The data is publicly available \dataRepo{}.

\vspace{1ex}
\sloppy
\noindent\textbf{Acknowledgments.~}Igor is supported by the FACEPE
fellowship IBPG-0123-1.03/17. This research was partially funded by
INES 2.0, FACEPE grants PRONEX APQ 0388-1.03/14 and APQ-0399-1.03/17,
and CNPq grant 465614/2014-0.

\balance
\bibliographystyle{spmpsci}
\bibliography{references,tmp}

\end{document}